\documentclass[aps,pra,twocolumn,showpacs,superscriptaddress,amsmath,amssymb,floatfix]{revtex4-1}
\usepackage{epsfig}
\usepackage{graphicx}
\usepackage{bm}
\usepackage{subfigure}

\def\be{\begin{equation}}
\def\ee{\end{equation}}
\def\bea{\begin{eqnarray}}
\def\eea{\end{eqnarray}}

\begin{document}
\title{An operational approach to indirectly measuring tunneling time}

\author{Yunjin Choi}
\affiliation{Department of Physics and Astronomy \& Rochester Theory Center, University of Rochester, Rochester, New York 14627, USA}

\author{Andrew N. Jordan}

\affiliation{Department of Physics and Astronomy \& Rochester Theory Center, University of Rochester, Rochester, New York 14627, USA}

\affiliation{Institute of Quantum Studies, Chapman University, 1 University Drive, Orange, CA 92866, USA}

\date{\today}

\begin{abstract}
The tunneling time through an arbitrary bounded one-dimensional barrier is investigated using the dwell time operator.
We relate the tunneling time to the conditioned average of the dwell time operator because of the natural post-selection in the case of successful tunneling.
We discuss an indirect measurement by timing the particle, and show we are able to reconstruct the conditioned average value of the dwell time operator by applying the contextual values formalism for generalized measurements based on the physics of Larmor precession.
The experimentally measurable tunneling time in the weak interaction limit is given by the weak value of the dwell time operator plus a measurement-context dependent disturbance term. We show how the expectation value and higher moments of the dwell time operator can be extracted from measurement data of the particle's spin.
\end{abstract}

\maketitle

\section{Introduction}\label{introduction}
One of the oldest, unsettled problems of quantum mechanics is that of the
tunneling time \cite{Condon,MacColl}. The seemingly simple question is how long does a tunneling
particle stay in the classically forbidden region. The question has generated continuous interest both in theory and experiment and still remains controversial \cite{Buttier.Landauer,Sokolovski.Baskin,Hauge.Stovneng,Steinberg0,Landauer.Martin,Winful,Davies,Aharonov.Erez,Sokolovski,Ordonez,Boyd}.
One source of difficulty
in dealing with this problem is the fact there is no generally accepted
time operator in quantum mechanics that obeys a canonical commutation
relation with the Hamiltonian. This was shown to be impossible for bound systems by Pauli because such a time operator would generate energy translations lower
than the ground state \cite{Pauli}. Numerous works have defined non-standard time
operators that work around the Pauli objection \cite{Kobe1,Kobe2} such as restricted non-self-adjoint time operators \cite{Werner} or generalized
measurement \cite{Leon}. In this work we are concerned with another kind of time
operator, which measures time differences, rather than absolute time \cite{Jaworski.Wardlaw}. Such
a dwell time operator commutes with the Hamiltonian and represents how long
a particle spends in a region of space \cite{Muga2}. We argue this operator is more
appropriate to the question of tunneling time. The issue remains, however,
how such an operator can be measured in an operational way in a laboratory: If one
prepares an incident wave packet, how does one know when to start the
clock, or how to stop it?

There have been direct methods proposed for measuring the tunneling time for a particle prepared in a Gaussian wave packet.  In this situation, it is the group delay, or shift of the peak of the wave packet that is used.
However, such an approach has been criticized because the tunneling time is defined as the arrival time of the peak of the wave packet after the barrier \cite{Winful,Buttiker.Washburn}. When this definition is used to calculate traversal velocities as the ratio of the length of the barrier to the tunneling time, it leads to the Hartman effect which exhibits superluminal velocities \cite{Hartman}. This draw back leads us to consider other definitions. An ingenious indirect way to measure the tunneling time was proposed by B\"{u}ttiker \cite{Buttiker} by refining Rybachenko's Larmor clock idea \cite{Baz,Rybachenko}. The idea is to attach a stopwatch to the particle that would turn on when the particle
was in the tunneling region, and turn off again when it emerged. This is
accomplished by using a particle with spin, and the physics of Larmor
precession: by applying a small magnetic field only in the classically
forbidden region, the spin will precess, and the ratio of the subtended
spin angle to the Larmor frequency defines the tunneling time, $\tau_y$.
Despite the conceptual clarity of this idea, the fact the spin
experiences different barrier heights depending on its orientation leads
also to spin rotation in a direction perpendicular to the precession plane,
which when divided by the Larmor frequency gives another time, $\tau_z$.
This effect provides an interpretational difficulty of which angle to use (if
this is indeed the correct procedure). B\"{u}ttiker suggested using a combination of both
times, $\sqrt{\tau_y^2 + \tau_z^2}$, while others advocated both times being
used separately as time scales, despite the fact $\tau_z$ can be negative \cite{Landauer.Martin}. While these times are intuitive, the argument is heuristic since they are not derived from an operator.

Steinberg stressed the fact that the tunneling time can only be defined
for the particles that actually tunnel through the barrier, and
consequently this definition only applies to a small fraction of all
particles in the system that are naturally post-selected \cite{Steinberg}. Such a
post-selected average can be calculated as a weak value of a time operator \cite{AAV,AV,Aharonov1,Aharonov2,Aharonov3,dressel0}, where the spin functions as a meter. Steinberg considered a projection operator on the
tunneling region scaled by the inverse particle current as the system operator. He found that the Larmor times $\tau_y$ and $\tau_z$ could be
understood as the real and imaginary parts of the weak value expression.

The connection between the results of a generalized measurement and the
measured operator is quite subtle. To further understand this interplay,
Dressel \textit{et al}. \cite{dressel0,dressel1} proposed the use of generalized eigenvalues of an operator,
called contextual values (CVs), which would be weighted with the frequencies of
detector outputs in order to calculate averages and moments of system
operator.

The purpose of the present paper is to reconsider the Larmor clock system which introduces a natural way to approach the tunneling problem. However, we deal with it by introducing a Hermitian observable for the time operator.
We consider the Larmor clock system as a generalized detector and use the Larmor times as detector outputs, rather than as system tunneling times, that will enable us to reconstruct the average and higher moments of the dwell time operator.
This formalism also allows us to define the tunneling time as the conditioned average of the
dwell time operator, and we find results which are related, but not
identical to, the weak value results of Steinberg \cite{Steinberg}. The discrepancy originates from the use of a different starting operator, and the
noncommutativity of the unitary part of the measurement operator with the
post-selection when the full dwell time operator is considered \cite{dressel2}. This approach also gives a pragmatic prescription for experimental implementation of this idea.

The rest of the paper is organized as follows. In Sec. \ref{tunneling_time_measurement}, we introduce some elementary properties of the one-dimensional scattering system, the dwell time operator and the weak value. In Sec. \ref{Larmor_system}, we analyze the Larmor system based on the generalized measurement and CV approach in detail. We discuss the spin rotated by the interaction in Sec. \ref{Spin_rotation}, and the detailed results of CVs for different shapes of the potential barrier in Sec. \ref{Decomposition_and_measurement}. The tunneling time defined as a conditioned average of the dwell time operator is shown in Sec. \ref{conditioned average of the dwell time operator}, and we provide the second moment of the dwell time operator in Sec. \ref{Second_moment}. We compare Steinberg's approach \cite{Steinberg,Steinberg2} with ours in Sec. \ref{Comparison_with_another_idea}. Finally, we give our conclusions in Sec.\ref{Conclusion}.

\section{Tunnelling time measurement}\label{tunneling_time_measurement}
Let us consider a particle of mass $m$ with energy $E=p^2/2m$ in a one-dimensional system with the spatial coordinate $x$ and a potential barrier centered at $x=0$, $V(x)$. In the position representation, the complete basis of scattering stationary states of the Hamiltonian
\begin{equation}
\hat{H}=\hat{p}^2/2m+V(x)\Theta_B(x),\label{Hamiltonian_spinless}
\end{equation}
have the following forms for the left (l) and right (r) incoming states, $k > 0$
\begin{eqnarray*}
\langle x|\phi_{l}(k)\rangle=\frac{1}{\sqrt{2\pi}}\left\{
                             \begin{array}{ll}
                             e^{ikx}+r^l(k) e^{-ikx}, & x < -d/2 \\
                             t(k) e^{ikx}, & x > d/2
                             \end{array}
                             \right.,\nonumber\\
\langle x|\phi_{r}(k)\rangle=\frac{1}{\sqrt{2\pi}}\left\{
                             \begin{array}{ll}
                             t(k)e^{-ikx} , & x < -d/2 \\
                             e^{-ikx}+r^r(k) e^{ikx}, & x > d/2
                             \end{array}
                             \right.,\nonumber
\end{eqnarray*}
where $\Theta_B(x)$ takes the value $1$ in the barrier region $[-d/2,d/2]$ and is zero elsewhere. The coefficients, $t(k)$ and $r^{l/r}(k)$, are the transmission and reflection amplitudes of left/right coming states, respectively. Since the reflection phases can depend on the incident direction, we keep the labels $l/r$. We omit any explicit expression for the states in the interval $[-d/2,d/2]$ because it depends on the details of the potential.

\begin{figure}
    \includegraphics[width=7cm]{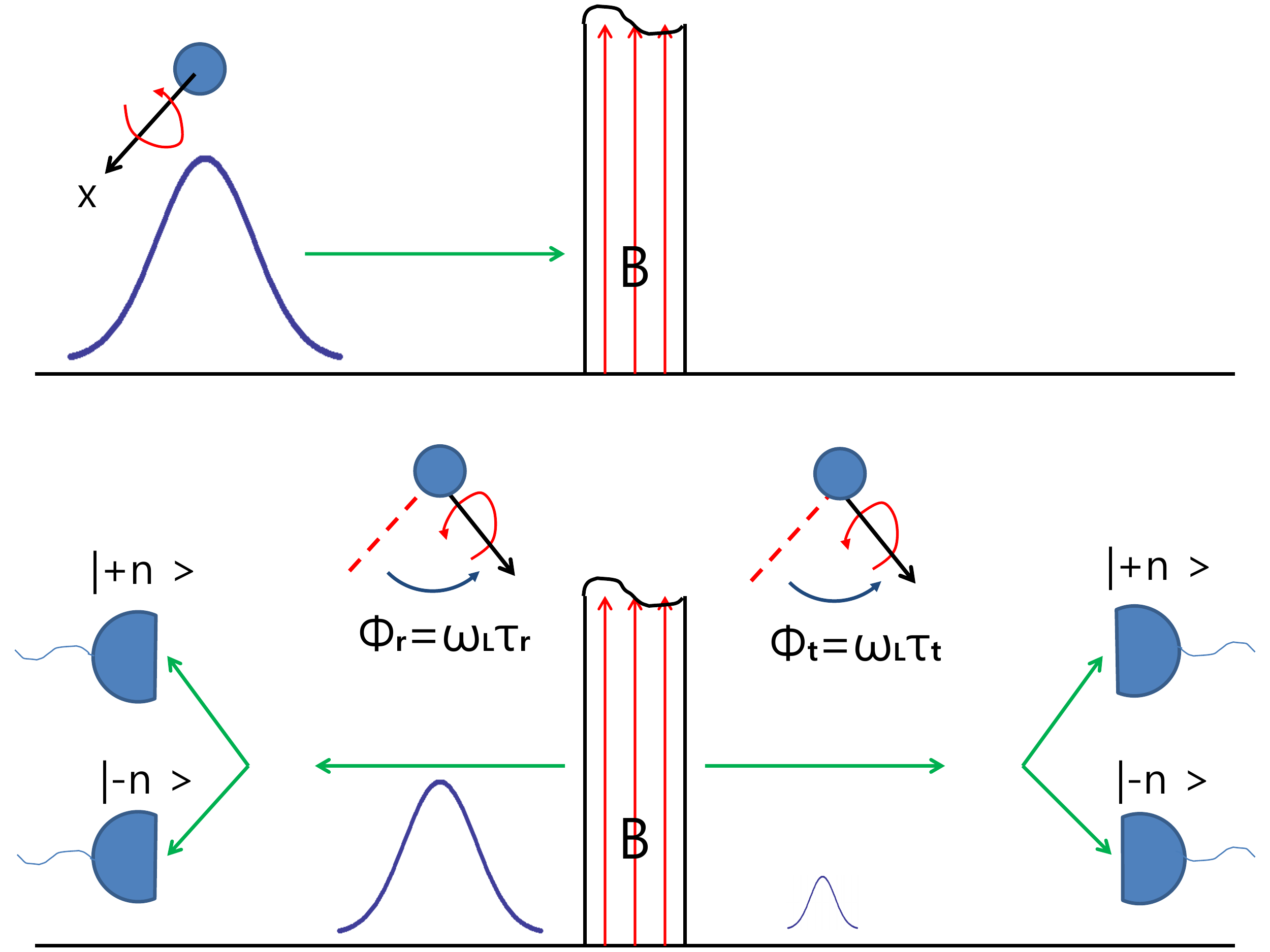}
    \caption{The one dimensional tunneling configuration of our system: A wave packet is traveling toward the potential barrier from the left hand side. After interaction with the barrier, the spin measurement is performed on the reflected and transmitted portion.}
    \label{fig1}
\end{figure}
When we consider the scattering process, there is a dilemma: On the one hand, in the tunneling problem (before interaction with the potential barrier) the particle is free.
On the other hand, the free particle eigenfunctions extend over all space with uniform density, so there is always an interaction with the potential barrier. To avoid this difficulty, we use a wave packet to describe the particle.
An initial wave packet prepared at $t_0\rightarrow-\infty$, $|\Psi_{in}(t_0\rightarrow-\infty)\rangle$ on the left hand side of the barrier, propagates toward the barrier, taking time $t$ before interaction with the barrier. We write the initial state by using the left coming positive momentum state $|k\rangle$, $k>0$, as
\begin{equation}
|\Psi_{in}(t)\rangle=\int dk A(k)|k\rangle e^{-i\omega(k) (t-t_0)}, \label{initialstate}
\end{equation}
where $\omega(k)=\hbar k^2/2m$, and $A(k)$ is a sharply peaked momentum distribution at mean momentum $k_c$ and insures only positive momentum contribution to the integral. Moreover, we assume the condition $k_c < k_0(x)$, where $k_0(x)=\sqrt{2mV(x)}/\hbar$, to make sure the particle is in the tunneling regime.

\subsection{Dwell time operator}
The dwell time operator for a particle staying in a region of interest $[-d/2,d/2]$ is defined as
\begin{equation}
\hat{T}_D=\int_{-\infty}^{\infty}dt e^{i\hat{H}t/\hbar}\int_{-d/2}^{d/2} dx|x\rangle \langle x| e^{-i\hat{H}t/\hbar},\label{def_dwelltime_operator}
\end{equation}
with $\hat{H}$, the system Hamiltonian \cite{Jaworski.Wardlaw,Muga2}. For the initial wave packet (\ref{initialstate}), the expectation value of this operator gives a momentum averaged dwell time,
\begin{eqnarray}
\tau_D  &=& \int_{-\infty}^{\infty}dt\int_{-d/2}^{d/2}dx |\Psi_{in}(x,t)|^2\nonumber\\
&=& \int_{0}^{\infty} dk |A(k)|^2\tau_d(k),
\end{eqnarray}
where $\tau_d(k)$ is the dwell time which is defined within the context of a stationary state scattering problem as the average number of particles within a region, $[-d/2,d/2]$, divided by the average number entering the region per unit time,
\begin{equation}
\tau_d(k)\equiv\frac{1}{j_{in}} \int_{-d/2}^{d/2} |\langle x|\phi_l(k)\rangle|^2 dx.\label{dwelltime(k)}
\end{equation}
Here $j_{in}=\hbar k/m$ is the incoming probability current density and $|\phi_l(k)\rangle$ is the left coming scattering state of the time independent Schr\"{o}dinger equation \cite{Buttiker}. Thus we see the total dwell time expectation is simply the weighted average of $\tau_d(k)$ over all $k$.
Since the dwell time operator commutes with the Hamiltonian $[\hat{T}_D,\hat{H}]=0$ \cite{Muga2} (see Appendix A), the dwell time operator can be expressed in the scattering basis,
\begin{eqnarray}
\hat{T}_D= \int_{0}^{\infty} dk \sum_{i,i'=r,l}C_{i,i'}(k)|\phi_i(k)\rangle\langle \phi_{i'}(k)|, \label{dwelltimeoperator}
\end{eqnarray}
by defining four elements of the operator, $C_{i,i'}(k)=\frac{m}{\hbar k} \int_{-d/2}^{d/2}dx\langle \phi_{i}(k)|x\rangle\langle x|\phi_{i'}(k)\rangle$ with $\{i,i'\}\in \{r,l \}$ for right $(r)$ and left $(l)$ coming states, respectively. The off diagonal elements of the matrix in the scattering basis $\{|\phi_l(k)\rangle, |\phi_r(k)\rangle\}$ are $C_{r,l}(k)=C_{l,r}^*(k)$ which confirm the operator is Hermitian.  When $i'=i$, the element $C_{i,i}(k)=\tau_d(k)$ is the same as the definition of the dwell time in Eq. (\ref{dwelltime(k)}) but with incoming state from right or left ($i=r,l$).

We wish to pick a specific wavenumber $k$ and measure the expectation value of the dwell time operator for that wavenumber, so we define the $k$ dependent dwell time operator
\begin{eqnarray}
\hat{T}_d(k)\equiv \sum_{i,i'=r,l}C_{i,i'}(k)|\phi_i(k)\rangle\langle \phi_{i'}(k)|. \label{dwelltimeoperator_k}
\end{eqnarray}
In what follows, we will focus on $\hat{T}_d(k)$ rather than $\hat{T}_D$. We will see in the next section why $k$ dependent dwell time operator is more relevant for us.

\subsection{Weak value}
The weak value, as a result of weakly measuring a operator $\hat{A}$, is assigned if a system is preselected in an initial state $|\psi_{i}\rangle$ and post-selected on a final state $|\psi_f\rangle$.
To illustrate this, a system operator $\hat{A}$ is weakly coupled to a detector momentum operator $\hat{p}$, and its time-dependent interaction Hamiltonian is
\begin{equation}
\hat{H}_{int}(t)=g(t)\hat{A}\otimes\hat{p}, \label{int_hamil_general}
\end{equation}
where the interaction profile $g(t)$ gives an effective coupling parameter $g=\int_{0}^T dt g(t)$ over the time interval $T$. Then the result of measuring the operator $\hat{A}$, subject to the pre and post-selection, is given by
\begin{eqnarray}
A^w=\frac{\langle\psi_f|\hat{A}|\psi_i\rangle}{\langle \psi_f|\psi_i\rangle}.\label{weak_value_general}
\end{eqnarray}

Now, we want to consider a wave packet tunneling through the potential barrier and calculate how long it takes to do so, the tunneling time. In this case, the weak value expression of the dwell time operator intuitively gives the solution to this problem, since the tunneling time is defined by the state which is initially prepared on the left hand side of the barrier (for a case of left coming wave packet), but later on is found at the right hand side. We consider the initial state $|\psi_i\rangle=\lim_{\tau\rightarrow-\infty}|k(\tau)\rangle$, which describes the right moving free particles on the left region before interaction with the potential barrier, as well as a post-selected state $|\psi_f\rangle=\lim_{\tau\rightarrow\infty}|k(\tau)\rangle$ describing a particle moving to the right on the right hand side of the barrier.
In this case, the weak value of the dwell time operator $\hat{T}_d(k)$ is calculated to be
\begin{eqnarray}
T_d^w(k)=C_{l,l}+\frac{r^r}{t}C_{r,l}\label{weakvalue_formalism},
\end{eqnarray}
which involves the sum of dwell time with the off-diagonal element of dwell time operator amplified by $r^r/t$.


To gain deeper insight into this result, the contextual value (CV) formalism \cite{dressel0,dressel1} puts forward a more general starting point and the weak value is obtained as a special case in this formalism. To more deeply understand how this pre- and post- selected quantity can be measured, we use the CV formalism to construct the result that can be measured in the lab. This approach gives a conceptually clear method for constructing the conditional average of the dwell time operator. To measure the tunneling time indirectly by using the physical detector, we use the Larmor clock \cite{Buttiker} which is a particular realization of a von Neumann-style interaction \cite{vonNeumann} for a meter corresponding to a spin $1/2$ particle. The Larmor system is defined in the weak coupling regime, so we can compare the measurable tunneling time to the weak value.

\section{Larmor system and generalized measurement}\label{Larmor_system}
The Larmor clock measurement scheme is when a small uniform magnetic field pointing in $z$ direction, $\hat{B}=B_0\hat{z}$, is confined to the barrier region. Suppose that the quantum particle has spin $\hbar/2$ and is initially polarized in the $+x$ direction in the incident beam. To measure the dwell time operator inside the tunneling region, we consider the spin as our detector. The Hamiltonian of the Larmor system is then given by
\begin{equation}
H_L=\frac{\hat{p}^2}{2m}+(V(x)-\frac{\hbar}{2}\omega_L\hat{\sigma}_z)\Theta_B(x),
\end{equation}
where $\omega_L=g\mu B_0/\hbar$ is the Larmor frequency, $g$ is the gyromagnetic ratio, and $\mu$ is the absolute value of the magnetic moment. The stationary solution of this Hamiltonian is a combination of two plane waves with spin components. Again, we only consider the incoming scattering stationary states, which are a complete basis with the following position representation. The spinful scattering states generalize to $|s,\phi_{l/r,\pm z}(k)\rangle=|\pm z\rangle |\phi_{l/r,\pm z}(k)\rangle$,
where the scattering states of the left and right coming state for the spin $\pm z$ in the position, $x$, domain are
\begin{eqnarray}
\langle x|\phi_{l,\pm z}(k)\rangle =\frac{1}{\sqrt{2\pi}}\left\{
                             \begin{array}{ll}
                             e^{ikx}+r_{\pm}^l(k) e^{-ikx}, & x < -d/2 \\
                             t_{\pm}^l(k) e^{ikx}, & x > d/2
                             \end{array}
                             \right.,\nonumber\\
\langle x|\phi_{r,\pm z}(k)\rangle =\frac{1}{\sqrt{2\pi}} \left\{
                             \begin{array}{lll}
                             t_{\pm}^r(k) e^{-ikx}, & x < -d/2 \\
                             e^{-ikx}+r_{\pm}^r(k) e^{ikx}, & x > d/2
                             \end{array}
                             \right.,  \label{scatteringstates}
\end{eqnarray}
where $|\pm z\rangle$ are the eigenstates of $\hat{\sigma}_z$. They are delta function normalized, $\langle s,\phi_{i,m}(k)|s,\phi_{i',m'}(k')\rangle=\delta_{i,i'}\delta_{m,m'}\delta(k-k')$. The transmission and reflection probabilities of left or right coming particles for spin $\pm z$ components are defined as $t_{\pm}^{l/r}, r_{\pm}^{l/r}$. For the spinless case, $t_{\pm}^{l/r}(k)$ and $r_{\pm}^{l/r}(k)$ merge to $t^{l/r}(k)$ and $r^{l/r}(k)$, and only two eigenstates, $|\phi_{l}(k)\rangle, |\phi_{r}(k)\rangle$, exist.


Even though the tunneling problem is most naturally treated as a time dependent wave packet traversing the barrier, it is possible nevertheless to develop a stationary approach so the scattering problem can be expressed in terms of the scattering amplitude at a given energy and we simply focus on the stationary state solution at momentum $k$.

We, therefore, consider the free particle state vector of the Hamiltonian, $\hat{H}_0=p^2/2m$, $|\pm k(\tau)\rangle=e^{-(i/\hbar)\hat{H}_0(\tau-\tau_0)}|k(\tau_0)\rangle$, for $k>0$ where $\tau_0$ is the initial time.
To describe the scattering experiment from the potential barrier, we take our initial incoming state vector to be $|k(\tau\rightarrow-\infty)\rangle$ which never experiences the potential barrier and the state we post-select on after scattering to be $|k(\tau\rightarrow\infty)\rangle$.
The connection between pre- and post-selection states is given by the scattering matrix (S-matrix) $\hat{S}$ which contains the effect of interaction.

Since the collision conserves energy, the S-matrix only connects momentum states with the same energy. For the $1$D scattering case (including the spin degree of freedom), we can represent the S-matrix for the momentum $k$ as $2\times 2$ matrices,
\begin{eqnarray}
\hat{S}_k
&=&\left(
          \begin{array}{cc}
            t_+ & r_+^r \\
            r_+^l & t_+ \\
          \end{array}
        \right)|+z\rangle\langle+z|
        + \left(
          \begin{array}{cc}
            t_- & r_-^r \\
            r_-^l & t_- \\
          \end{array}
        \right)|-z\rangle\langle-z|,\nonumber\\
\end{eqnarray}
where the $2\times 2$ matrices are in the basis of left and right movers, $(\substack{1\\ 0})=|k\rangle$ and $(\substack{0\\1})=|-k\rangle$, respectively.

Now, we specialize the results by an approximation based on the required condition of the Larmor system: the applied magnetic field is small so the change in potential energy created by this field is very small compared to both the height of the barrier and the difference between the height of the barrier and the kinetic energy. This condition additionally gives weak interaction between the system and the detector. The weakness condition allows us to approximate the transmission and reflection amplitudes to first order in $\omega_L$
\begin{eqnarray}
t_{\pm}&\simeq& t\: \left[1\pm\frac{1}{2}\omega_L (\tau_{zt}+i\tau_{yt})\right], \nonumber\\
r_{\pm}^{l/r}&\simeq& r^{l/r} \left[ 1\pm\frac{1}{2}\omega_L(\tau_{zr}+i\tau_{yr}^{l/r})\right], \label{approximation}
\end{eqnarray}
where the times $\tau_{(z/y)(t/r)}$ are defined as the derivatives of the transmitted or reflected probability of the barrier, and the derivative of its phase. For example, if we consider the box potential $V(x)\Theta_B(x)=V_0\Theta_B(x)$, the expressions of $\tau_{zt/zr}$ and $\tau_{yt/yr}$ are
\begin{eqnarray}
\tau_{zt}&=&-(m/\hbar\kappa)\partial \ln T^{\frac{1}{2}}/\partial\kappa,\nonumber \\
\tau_{zr}&=&-(T/R)\tau_{zt},\nonumber\\
\tau_{yt}&=&-(m/\hbar\kappa)\partial\varphi_t/\partial\kappa,\label{larmor_times}\\
\tau_{yr}^{l/r}&=&-(m/\hbar\kappa)\partial\varphi_r^{l/r}/\partial\kappa, \nonumber
\end{eqnarray}
where $\kappa=\sqrt{k_0^2-k^2}$ for $k_0=\sqrt{2mV_0}/\hbar$. The amplitude related times $\tau_{zt/zr}$ come from the logarithmic derivative of the transmitted and reflected probabilities ($T$ and $R$), and $\tau_{yt/yr}^{r/l}$ are phase ($\varphi_{t/r}$) related times. The times (\ref{larmor_times}) are the Larmor times that B\"{u}ttiker defined to represent the tunneling time \cite{Buttiker}. The left and right coming dependence of the amplitudes only belong to the phase of the reflected amplitude, $\varphi_{r}^{l/r}$ because the transmitted phase is the same for both directions. Thus the superscript $l/r$ of the times only appears in $\tau_{yr}^{l/r}$. A more detailed discussion of Eq. (\ref{larmor_times}) for the arbitrary potential case is given in Refs. \cite{Pollak,Gasparian}.

Now, we suppose that the initially prepared joint state of the system and detector is a product state, $\lim_{\tau\rightarrow-\infty}|k(\tau)\rangle|+x\rangle$. As time goes on, the joint state evolves under a unitary evolution which contains the system-detector interaction: the interaction of the system state both with the barrier $V(x)$ and the spin through the applied infinitesimal magnetic field. As $\tau\rightarrow\infty$, the system will again be in scattering states given from the $\tau\rightarrow-\infty$ scattering states.
The unitary interaction will, then, entangle the system with the detector so that performing a direct measurement on the detector (spin system) will lead to an indirect measurement being performed on the system.

Compared to a von-Neumann interaction Eq. (\ref{int_hamil_general}), $\hat{\sigma}_z$ plays the role of detector $\hat{p}$. Consequently, the azimuthal angle of this axis, which is the phase difference between spin $\pm z$ components, will be the conjugate pointer position \cite{Dirac}. Therefore, spin measurement for an arbitrary direction is analogous to measuring some combination of the free particle detector's position and momentum degree of freedom. We measure, after the barrier, the spin post-selecting on the $\pm n$ direction for both the transmitted and the reflected particles.
The two orthonormal spin states (experimentally chosen) are
\begin{eqnarray}
|+n\rangle&=&\cos\frac{\theta}{2}|+z\rangle+e^{i\phi}\sin\frac{\theta}{2}|-z\rangle,\nonumber\\
|-n\rangle&=&\sin\frac{\theta}{2}|+z\rangle+e^{i (\phi+\pi)}\cos\frac{\theta}{2}|-z\rangle,\label{arb_spin}
\end{eqnarray}
with $0 < \theta <\pi$ and $0<\phi<2 \pi$.

Since the system and detector states are entangled, a measurement on a particular detector spin $|\pm n\rangle$ is equivalent to the measurement operator $\hat{M}_{m}$ on the system,
\begin{eqnarray}
\hat{M}_m&\equiv&\langle m|\hat{S}_k|+x\rangle\nonumber\\
&\simeq&\left[  \left(
             \begin{array}{cc}
               t & r^r \\
               r^l & t \\
             \end{array}
           \right)\langle m|+x\rangle\right.\nonumber\\
  &&\left.+\frac{\omega_L}{2} \left(
             \begin{array}{cc}
               t(\tau_{zt}+i\tau_{yt}) & r^r(\tau_{zr}+i\tau_{yr}^r) \\
               r^l(\tau_{zr}+i\tau_{yr}^l) & t(\tau_{zt}+i\tau_{yt}) \\
             \end{array}
           \right)\langle m|-x\rangle
  \right]\nonumber\\
&\equiv&\hat{M}_m^{0}+\omega_L\hat{M}_m^{(1)}\label{measurement_operator},
\end{eqnarray}
where $m=\pm n$. The measurement of the spin state is simultaneously accompanied by the measurement of the position or momentum of the particle, since our momentum states, $|\pm k(\tau\rightarrow\infty)\rangle$ contain the left/right position information. Defining the following momentum projection operators
\begin{eqnarray}
\hat{\Pi}_r&\equiv&\lim_{\tau\rightarrow\infty}|k(\tau)\rangle\langle k(\tau)|=\left(
                                                \begin{array}{cc}
                                                  1 & 0 \\
                                                  0 & 0 \\
                                                \end{array}
                                              \right),\nonumber\\
\hat{\Pi}_l&\equiv&\lim_{\tau\rightarrow\infty}|-k(\tau)\rangle\langle -k(\tau)| =\left(
                                                \begin{array}{cc}
                                                  0 & 0 \\
                                                  0 & 1 \\
                                                \end{array}
                                              \right),\label{projection_operator}
\end{eqnarray}
permits us to define our measurement operator $\hat{\mathcal{M}}_{p,m}\equiv\hat{\Pi}_p\hat{M}_m$ where $p=\{r,l\}$ and $m=\{+n,-n\}$.
For later convenience, we introduce two values, given by the overlap of spin states,
\begin{eqnarray}
x^{0}_{\pm n}&=&|\langle \pm n|+x\rangle|^2,\\
x^{(1)}_{n}&=&\langle +x|+n\rangle\langle+n|-x\rangle=-\langle +x|-n\rangle\langle-n|-x\rangle,\nonumber
\end{eqnarray}
and the complex times,
\begin{eqnarray}
\tau_t&\equiv&\tau_{zt}+i\tau_{yt},\nonumber\\
\tau_r^r&\equiv&\tau_{zr}+i\tau_{yr}^r,\label{complextimes}\\
\tau_{r}^l&\equiv&\tau_{zr}+i\tau_{yr}^l.\nonumber
\end{eqnarray}
Then the probability operators (or POVM elements) on the system are defined in the momentum basis as
\begin{eqnarray}
\hat{E}_{r,\pm n}&=&\hat{\mathcal{M}}_{r,\pm n}^{\dag}\hat{\mathcal{M}}_{r,\pm n}\nonumber\\
&\simeq&\left(
     \begin{array}{cc}
       T & t^*r^r \\
       r^{r*}t & R \\
     \end{array}
   \right)x_{\pm n}^0
   \pm\frac{\omega_L}{2}\left[ \left(
                     \begin{array}{cc}
                       T\tau_{t} & t^*r^r\tau_{r}^r \\
                       r^{r*}t\tau_t & R\tau_r^r \\
                     \end{array}
                   \right)x_n^{(1)} \right.\nonumber\\
   &&+\left.  \left(
                     \begin{array}{cc}
                       T\tau_t^* & t^*r^r\tau_t^* \\
                       r^{r*}t\tau_r^{r*}& R\tau_r^{r*} \\
                     \end{array}
                   \right)x_n^{(1)*}
    \right],\label{povm_r}\\
\hat{E}_{l,\pm n}&=&\hat{\mathcal{M}}_{l,\pm n}^{\dag}\hat{\mathcal{M}}_{l,\pm n}\nonumber\\
&\simeq&\left(
     \begin{array}{cc}
       R & tr^{l*} \\
       r^{l}t^* & T \\
     \end{array}
   \right)x_{\pm n}^0
   \pm\frac{\omega_L}{2}\left[ \left(
                     \begin{array}{cc}
                       R\tau_r^l & r^{l*}t\tau_t \\
                       t^*r^{l}\tau_r^l & T\tau_t \\
                     \end{array}
                   \right)x_n^{(1)} \right.\nonumber\\
   &&+\left. \left(
                     \begin{array}{cc}
                       R\tau_r^{l*} & r^{l*}t\tau_r^{l*} \\
                       t^*r^{l}\tau_t^* & T\tau_t^* \\
                     \end{array}
                   \right)x_n^{(1)*}
    \right].\label{povm_l}
\end{eqnarray}
We note that the complex times Eq. (\ref{complextimes}) are just a book keeping device since the expectation value of the probability operator in any initial state, $\langle\hat{E}_{p,m}\rangle$, gives the real probability that the spin of the particle is measured in the $|m\rangle$ spin state on the $p$ side of the potential barrier (up to the first order of $\omega_L$).

\subsection{Spin rotated by the interaction}\label{Spin_rotation}
After the interaction of the particle with the potential barrier, the initial spin rotates differently whether it is transmitted or reflected. Not only the spin precession on the $x-y$ plane, but also rotation out of the plane is experienced by the spin state \cite{Buttiker}. The out of plane rotation is caused by the potential difference $V(x)\rightarrow V(x)\mp \hbar\omega_L/2$ from the effect of the magnetic field in the $\hat{\sigma}_z$ eigenbasis, leading to a difference of transmission amplitude. The spin state after the interaction can be obtained by the unitary $\hat{S}_k$ operating on the initial state. Taking the initial state to be $\lim_{\tau\rightarrow-\infty}|k(\tau)\rangle|+x\rangle$, the state after the interaction is
\begin{eqnarray}
&&\lim_{\tau\rightarrow-\infty}\hat{S}_k|k(\tau)\rangle|+x\rangle\nonumber\\
&&=\lim_{\tau\rightarrow\infty}\left[\frac{1}{\sqrt{2}}\left(
                                                                         \begin{array}{c}
                                                                           t_+ \\
                                                                           t_- \\
                                                                         \end{array}
                                                                       \right)|k(\tau)\rangle
                                                        +\frac{1}{\sqrt{2}}\left(
                                                                         \begin{array}{c}
                                                                           r_+^l \\
                                                                           r_-^l \\
                                                                         \end{array}
                                                                       \right)|-k(\tau)\rangle\right]\nonumber\\
&&=\lim_{\tau\rightarrow\infty}\left[\: t|k(\tau)\rangle|s_r\rangle  +r^l|-k(\tau)\rangle|s_l\rangle \:\right],
\end{eqnarray}
where we have defined the rotated spin state on the right $(r)$ or left $(l)$ hand side of the barrier
\begin{eqnarray}
|s_r\rangle&=&|+x\rangle+\frac{\omega_L}{2}(\tau_{zt}+i\tau_{yt})|-x\rangle,\label{s_r}\\
|s_l\rangle&=&|+x\rangle+\frac{\omega_L}{2}(\tau_{zr}+i\tau_{yr}^l)|-x\rangle,\label{s_l}
\end{eqnarray}
which are normalized states up to the first order of $\omega_L$. Note that the rotated states are naturally expressed with the complex times, Eq. (\ref{complextimes}). Therefore, when we post-select the system state on the right hand side of barrier, $|k(\tau\rightarrow\infty)\rangle$, the (re-normalized) rotated spin state will be Eq. (\ref{s_r}). It shows that the phase is changed by $\omega_L\tau_{yt}/2$ which causes the inplane precession, and the amplitude is changed by $\omega_L\tau_{zt}/2$, which causes out of plane rotation as are expected for the Larmor system \cite{Buttiker}.

\subsection{Decomposition and measurement of the dwell time operator with the probability operators}\label{Decomposition_and_measurement}
We are now in a position to discuss the measurement of the system operator $\hat{T}_d(k)$ given Eq. (\ref{dwelltimeoperator_k}) in more detail. The formalism of CV forms a bridge between the observable and the operations of the generalized measurement. The main idea of the CV formalism is that an observable can be completely measured indirectly using an imperfectly correlated detector by assigning an appropriate set of values to the detector outcomes. This approach gives the correct average value (expectation) of the operator for any initial state by construction and reproduces the generalized weak value formalism in the minimum disturbance limit and gives an operational way of computing conditional averages.
Therefore, to connect the system operator $\hat{T}_d(k)$ in Eq. (\ref{dwelltimeoperator_k}) with the probability operators $\hat{E}_{p,m}(k)$ in Eq. (\ref{povm_r}, \ref{povm_l}), we assign a set of CVs, $\{\alpha_{p,m}(k)\}$, for each outcome of the measurement,
\begin{eqnarray}
\hat{T}_d(k)&=&\alpha_{r,+n}(k)\hat{E}_{r,+n}(k)+\alpha_{r,-n}(k)\hat{E}_{r,-n}(k)\nonumber\\
&+&\alpha_{l,+n}(k)\hat{E}_{l,+n}(k)+\alpha_{l,-n}(k)\hat{E}_{l,-n}(k).
\label{cv_decomposition}
\end{eqnarray}
Although $\hat{T}_d(k)$ is given as a matrix in the scattering basis Eq. (\ref{dwelltimeoperator_k}), it has the same expression in the momentum basis because of the boundary conditions in the distance past and distant future:
\begin{eqnarray}
\lim_{\tau\rightarrow-\infty}\langle\phi_l(k,\tau)|k(\tau)\rangle&=&1,\nonumber\\
\lim_{\tau\rightarrow-\infty}\langle\phi_l(k,\tau)|-k(\tau)\rangle&=&0,\nonumber\\
\lim_{\tau\rightarrow-\infty}\langle\phi_r(k,\tau)|k(\tau)\rangle&=&0,\nonumber\\
\lim_{\tau\rightarrow-\infty}\langle\phi_r(k,\tau)|-k(\tau)\rangle&=&1,\nonumber\\
\lim_{\tau\rightarrow\infty}\langle\phi_l(k,\tau)|k(\tau)\rangle&=&t^*, \label{approx}\\
\lim_{\tau\rightarrow\infty}\langle\phi_l(k,\tau)|-k(\tau)\rangle&=&r^{l*},\nonumber\\
\lim_{\tau\rightarrow\infty}\langle\phi_r(k,\tau)|k(\tau)\rangle&=&r^{r*},\nonumber\\
\lim_{\tau\rightarrow\infty}\langle\phi_r(k,\tau)|-k(\tau)\rangle&=&t^*.\nonumber
\end{eqnarray}
To solve for the CVs in Eq. (\ref{cv_decomposition}) more easily, we linearly transform the dwell time operator and the probability operators by defining the unitarily transformed dwell time operator and probability operator as $\tilde{T}_d(k)$ and $\tilde{E}_{p,m}$, respectively,
\begin{eqnarray}
\tilde{T}_d(k)&\equiv&\hat{S}_k^0\:\hat{T}_d(k)\:\hat{S}_k^{0\dag}\label{matrix_trans}\\
&=&\left(
     \begin{array}{cc}
       T_{11} & T_{12} \\
       T_{21} & T_{22} \\
     \end{array}
   \right),\nonumber\\
\tilde{E}_{p,m}(k)&\equiv&\hat{S}_k^0\:\hat{E}_{p,m}(k)\:\hat{S}_k^{0\dag},\label{povm_tras}
\end{eqnarray}
where the unitary operator $\hat{S}_k^0$ is the S-matrix of the spinless system,
\begin{eqnarray}
\hat{S}_k^0=\left(
              \begin{array}{cc}
                t & r^r \\
                r^l & t \\
              \end{array}
            \right).
\end{eqnarray}
The detailed matrix elements of $\tilde{T}_d(k)$, given in terms of $C_{i,i'}$, are
\begin{eqnarray}
T_{11}&=&TC_{l,l}+RC_{r,r}+2\textmd{Re}[t^*r^rC_{rl}],\nonumber\\
T_{22}&=&RC_{l,l}+TC_{r,r}+2\textmd{Re}[r^{l*}tC_{rl}],\\
T_{12}&=&r^{l*}tC_{l,l}+r^{l*}r^rC_{r,l}+TC_{l,r}+t^*r^rC_{r,r},\nonumber\\
T_{21}&=&T_{12}^*,\nonumber
\end{eqnarray}
and the transformed probability operators are now
\begin{eqnarray}
\tilde{E}_{r,\pm n}&=&\left(
                      \begin{array}{cc}
                        1 & 0 \\
                        0 & 0 \\
                      \end{array}
                    \right)x_{\pm n}^0
                    \pm\omega_L \left(
                                  \begin{array}{cc}
                                    E_{r11} & E_{r12} \\
                                    E_{r12}^* & 0 \\
                                  \end{array}
                                \right),\label{transformed_er}\\
\tilde{E}_{l,\pm n}&=&\left(
                      \begin{array}{cc}
                        0 & 0 \\
                        0 & 1 \\
                      \end{array}
                    \right)x_{\pm n}^0
                    \pm\omega_L \left(
                                  \begin{array}{cc}
                                    0 & E_{l12} \\
                                    E_{l12}^* & E_{l22} \\
                                  \end{array}
                                \right).\label{transformed_el}
\end{eqnarray}
The newly introduced $\mathcal{O}(\omega_L)$ terms of Eq. (\ref{transformed_er}, \ref{transformed_el}) are easily calculated to be,
\begin{eqnarray}
E_{r11}&=& T\textmd{Re}[\tau_tx_n^{(1)}]+R\textmd{Re}[\tau_r^rx_n^{(1)}],\nonumber\\
E_{r12}&=&-r^{l*}t\frac{(\tau_r^l-\tau_t)^*}{2}x_n^{(1)},\nonumber\\
E_{l12}&=&r^{l*}t\frac{(\tau_r^l-\tau_t)^*}{2}x_n^{(1)*},\label{povm_new_elem}\\
E_{l22}&=&T\textmd{Re}[\tau_tx_n^{(1)}]+R\textmd{Re}[\tau_r^l x_n^{(1)}].\nonumber
\end{eqnarray}
Then the unitarily transformed Eq. (\ref{cv_decomposition}) with the same CVs gives,
\begin{eqnarray}
\tilde{T}_d(k)&=&\sum_{\substack{p=r,l\\m=\pm n}}\alpha_{p,m}\tilde{E}_{p,m}.\label{transformed_dwell}
\end{eqnarray}

For the weak measurement case, our CVs cannot have poles of greater order than $1/\omega_L$ \cite{dressel1}, and we make the ansatz the CVs may be expanded as $\alpha_{p,m}\simeq\alpha_{p,m}^0+\alpha_{p,m}^{(1)}/\omega_L$. It is convenient to define the differences and weighted sums of the CVs,
\begin{eqnarray}
\xi_{r/l}^{0}&\equiv&x_{+n}^{0}\:\alpha_{r/l,+n}^{0}+x_{-n}^{0}\:\alpha_{r/l,-n}^{0},\nonumber\\
\xi_{r/l}^{(1)}&\equiv&x_{+n}^{0}\:\alpha_{r/l,+n}^{(1)}+x_{-n}^{0}\:\alpha_{r/l,-n}^{(1)},\nonumber\\
\delta\alpha_{r/l}^{0}&\equiv&\alpha_{r/l,+n}^{0}-\alpha_{r/l,-n}^{0},\label{substitution}\\
\delta\alpha_{r/l}^{(1)}&\equiv&\alpha_{r/l,+n}^{(1)}-\alpha_{r/l,-n}^{(1)}.\nonumber
\end{eqnarray}
To solve Eq. (\ref{transformed_dwell}) with this ansatz, we can rewrite the matrix elements of $\tilde{T}_d(k)$, $T_{ii'}$, up to the first order of $\omega_L$ and $\omega_L^{-1}$
\begin{eqnarray}
T_{11}&=&\xi_{r}^{0}+E_{r11}\delta\alpha_{r}^{(1)}+\frac{1}{\omega_L}\xi_r^{(1)}+\omega_L\left[ E_{r11}\delta\alpha_r^{(0)} \right],\nonumber\\
T_{22}&=&\xi_{l}^{0}+E_{l22}\delta\alpha_l^{(1)}+\frac{1}{\omega_L}\xi_l^{(1)}+\omega_L\left[ E_{l22}\delta\alpha_l^{(0)} \right],\label{povm_new_rel}\\
T_{12}&=&E_{r12}\delta\alpha_r^{(1)}+E_{l12}\delta\alpha_{l}^{(1)}+\omega_L\left[ E_{r12}\delta\alpha_r^0+E_{l12}\delta\alpha_l^0 \right],\nonumber\\
T_{21}&=&T_{12}^*.\nonumber
\end{eqnarray}
For the weak interaction case, $\omega_L\rightarrow 0$, which is the case that we are considering, the $\mathcal{O}(\omega_L^{-1})$ terms should vanish to prevent a divergence in the weak limit. This indicates that two quantities must vanish,
\begin{eqnarray}
\xi_{r/l}^{(1)}&=&x_{+n}^{0}\:\alpha_{r/l,+n}^{(1)}+x_{-n}^{0}\:\alpha_{r/l,-n}^{(1)}=0.\label{xi}
\end{eqnarray}
Moreover, the elements of the dwell time operator do not depend on the measurement strength, so the $\mathcal{O}(\omega_L)$ terms should also vanish for finite $\omega_L$. This gives the condition
\begin{eqnarray}
\delta\alpha_{r/l}^{0}=\alpha_{r/l,+n}^0-\alpha_{r/l,-n}^0=0,\label{delta_alpha}
\end{eqnarray}
so $\alpha_{r,+n}^0=\alpha_{r,-n}^0$ and $\alpha_{l,+n}^0=\alpha_{l,-n}^0$. From Eqs. (\ref{xi}, \ref{delta_alpha}), we define
\begin{eqnarray}
\alpha_{r/l}^0&\equiv&\alpha_{r/l,+n}^0=\alpha_{r/l,-n}^0,\label{a_rl_0}\\
\alpha_{r/l}^{(1)}&\equiv&x_{+n}^0\:\alpha_{r/l,+n}^{(1)}=-x_{-n}^0\:\alpha_{r/l,-n}^{(1)},\label{a_rl_1}
\end{eqnarray}
and the CVs can be expressed in a simpler way,
\begin{eqnarray}
\alpha_{r/l,\pm n}&=&\alpha_{r/l}^0\pm\frac{1}{\omega_L x_{\pm n}^0}\alpha_{r/l}^{(1)}.\label{cv_rl}
\end{eqnarray}
The CVs start from the same spinless values, and then separate by the $\mathcal{O}(\omega_L^{-1})$ terms we defined as $\pm\alpha_{r/l}^{(1)}/x_{\pm n}^0$.

With these general considerations out of the way, we turn to the specific form of the CVs. The remaining nonzero conditions form $4$ equations for $4$ unknowns; we write it in matrix form:
\begin{eqnarray}
\left(
  \begin{array}{c}
    T_{11} \\
    T_{12} \\
    T_{21} \\
    T_{22} \\
  \end{array}
\right)=\left(
          \begin{array}{cccc}
            1 & 0 & E_{r11} & 0 \\
            0 & 0 & E_{r12} & E_{l12} \\
            0 & 0 & E_{r12}^* & E_{l12}^* \\
            0 & 1 & 0 & E_{l22} \\
          \end{array}
        \right)
        \left(
          \begin{array}{c}
            \xi_{r}^{0} \\
            \xi_{l}^{0} \\
            \delta\alpha_{r}^{(1)} \\
            \delta\alpha_{l}^{(1)} \\
          \end{array}
        \right).\label{matrix}
\end{eqnarray}
From Eqs. (\ref{a_rl_0}, \ref{a_rl_1}), we deduce $\xi_{r/l}^0=\alpha_{r/l}^0$ and $\delta\alpha_{r/l}^{(1)}=\alpha_{r/l}^{(1)}/x_{+n}^0x_{-n}^0$. Once we solve the matrix equation Eq. (\ref{matrix}), therefore, we produce the exact values of $\alpha_{r/l}^0$ and $\alpha_{r/l}^{(1)}$ in Eq. (\ref{cv_rl}).

Note from Eq. (\ref{matrix}) that when the spin post-selection direction $\hat{n}$ is in the $x-z$ plane of the spin Bloch sphere, $x_n^{(1)}$ is a real value and the $4\times 4$ matrix in Eq. (\ref{matrix}) does not have an inverse because the determinant of the $4\times4$ matrix is zero. In this case, there are no solutions to the four quantities $\xi_{r/l}^0$ and $\delta\alpha_{r/l}^{(1)}$. To see the physical reason for this, when we look at the diagonal elements of the probability operator of Eqs. (\ref{povm_r}, \ref{povm_l}), they contain only $\tau_{zt/zr}$, which is the amplitude deviation, and we lose the phase deviation $\tau_{yt/yr}$ in the $x-y$ plane.
In the other extreme, when we choose the spin polarization $\hat{n}$ in the $x-y$ plane, $x_n^{(1)}$ is purely imaginary and there are also no solutions for $\xi_{r/l}^0$ and $\delta\alpha_{r/l}^{(1)}$. This case deletes the $\tau_{zt/zr}$ information from the probability in Eqs. (\ref{povm_r}, \ref{povm_l}) and we lose the amplitude deviation from the interaction with the potential barrier. Therefore, to reconstruct the full dwell time operator in this measurement setup, we need to have all four times $\tau_{zt/zr}$ and $\tau_{yt/yr}$ at hand. Since the dwell time operator contains reflected and tunneled information, the complete solution of the CVs comes only when the related probability depends both on $\tau_{zt/zr}$ and $\tau_{yt/yr}$.

When the spin post-selection is neither in the $x-z$ plane or $x-y$ plane, we can invert the matrix Eq. (\ref{matrix}) and find the unique solutions of $\alpha_{r/l}^{0}$ and $\alpha_{r/l}^{(1)}$,
\begin{eqnarray}
\alpha_{r}^0&=&T_{11}+\textmd{Re}\left[ (T\tau_t+R\tau_r^r)x_n^{(1)} \right]\:f_r(n,\delta\tau),\nonumber\\
\alpha_l^0&=&T_{22}-\textmd{Re}\left[ (T\tau_t+R\tau_r^l)x_n^{(1)} \right]\:f_l(n,\delta\tau),\nonumber\\
\alpha_r^{(1)}&=&-x_{+n}^0x_{-n}^0\: f_r(n,\delta\tau),\label{cv_elements}\\
\alpha_l^{(1)}&=&x_{+n}^0x_{-n}^0\:f_l(n,\delta\tau),\nonumber
\end{eqnarray}
where we defined the functions
\begin{eqnarray}
f_r(n,\delta\tau)&=&\frac{\textmd{Im}\left[T_{12}\:t r^{r*}\:\delta\tau \:x_n^{(1)} \right]}{RT\:|\delta\tau|^2\:\textmd{Re}\left[x_n^{(1)}\right]\textmd{Im}\left[x_n^{(1)}\right]},\\
f_l(n,\delta\tau)&=&\frac{\textmd{Im}\left[T_{12}\:t^*r^l\:\delta\tau \:x_n^{(1)^*} \right]}{RT\:|\delta\tau|^2\:\textmd{Re}\left[x_n^{(1)}\right]\textmd{Im}\left[x_n^{(1)}\right]},
\end{eqnarray}
and $\delta\tau\equiv\tau_t-\tau_r^l=(\tau_t-\tau_r^r)^*$ is the difference of the complex times Eq. (\ref{complextimes}).
Results (\ref{cv_rl},\ref{cv_elements}) formally solve the CVs in the weak limit case.
To understand the results physically, let us consider the potential barrier for several cases.

\begin{figure}
\begin{center}
\subfigure{\includegraphics[width=7cm]{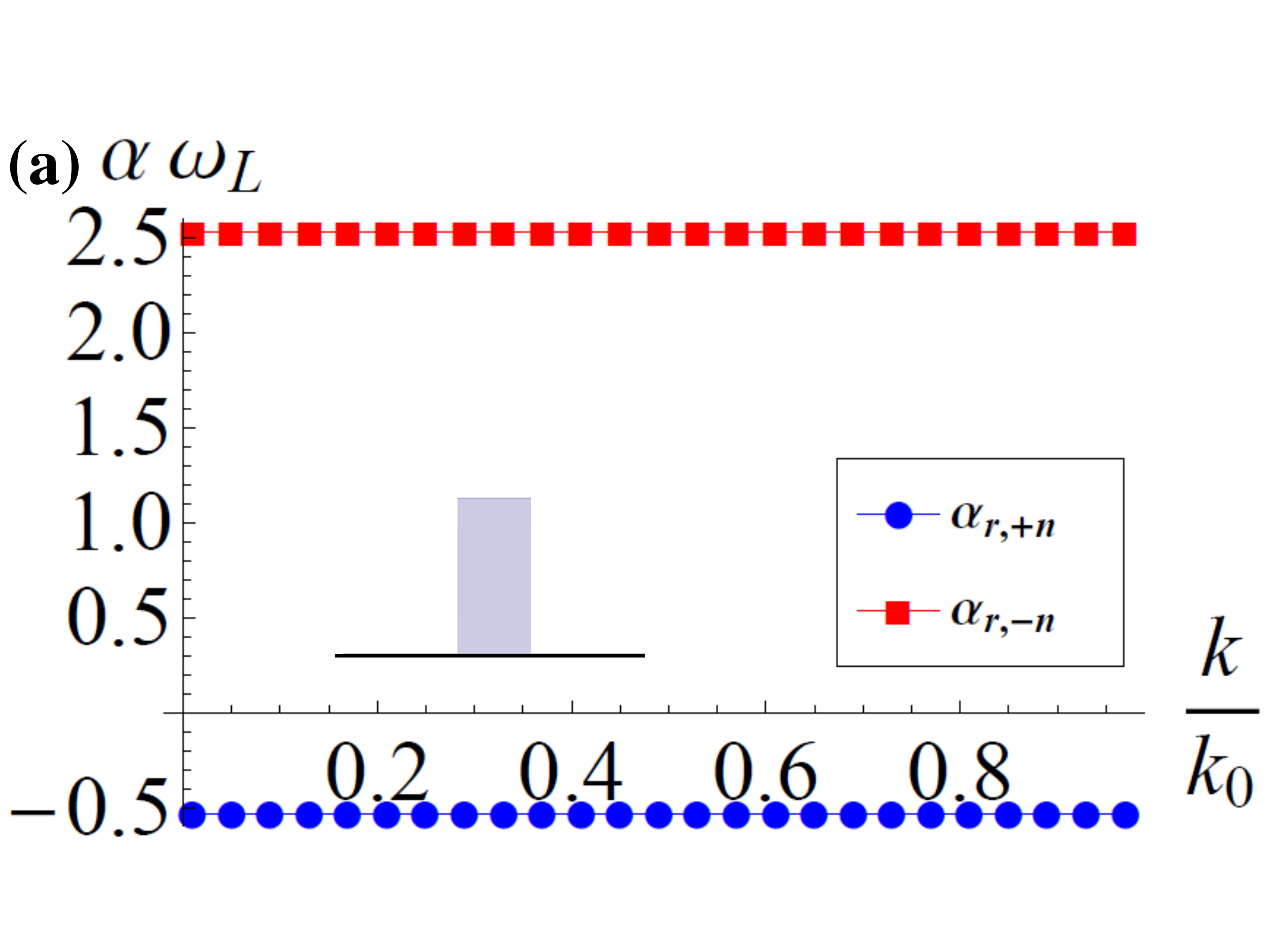}}
\subfigure{\includegraphics[width=7cm]{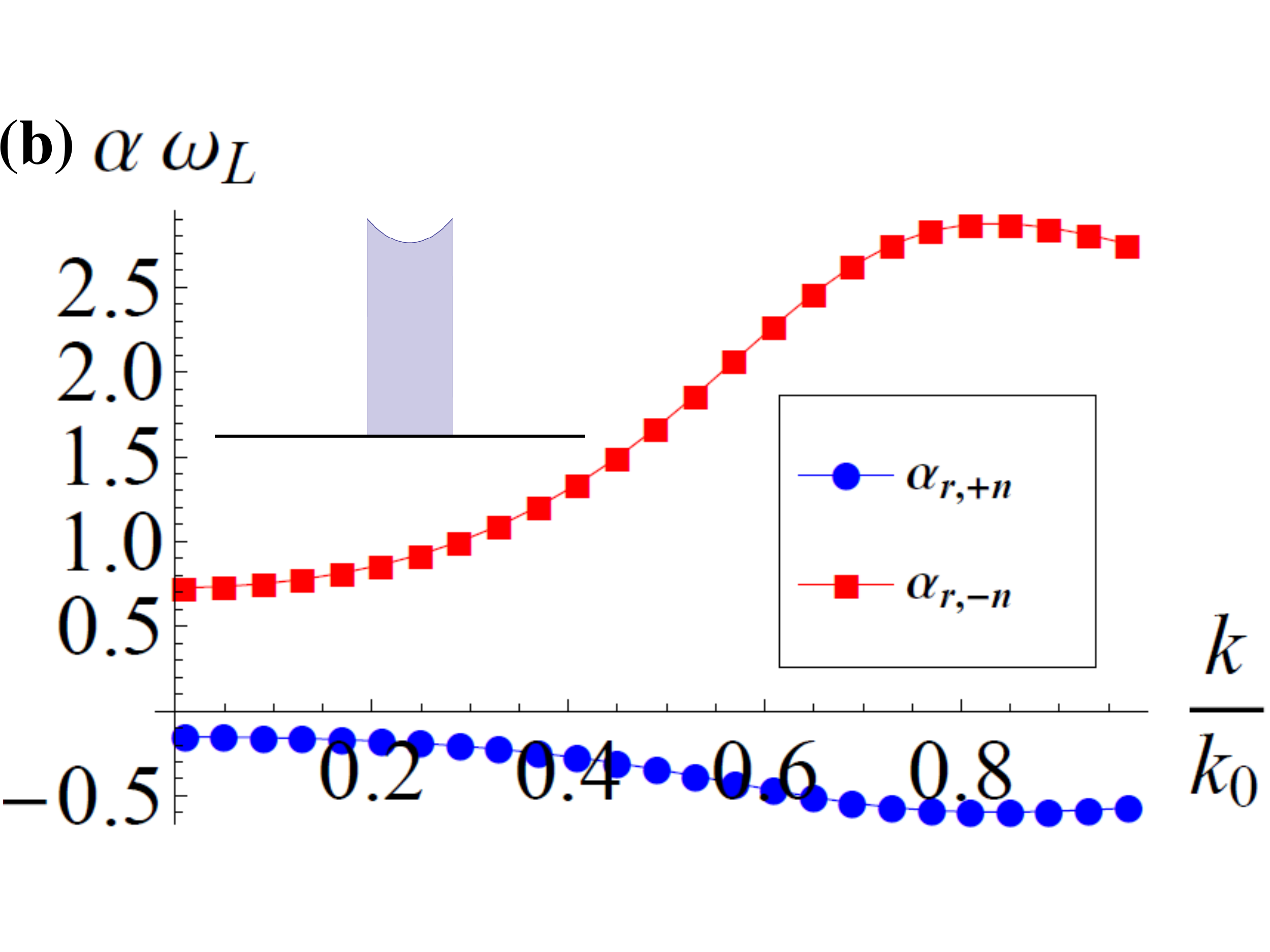}}
\subfigure{\includegraphics[width=7cm]{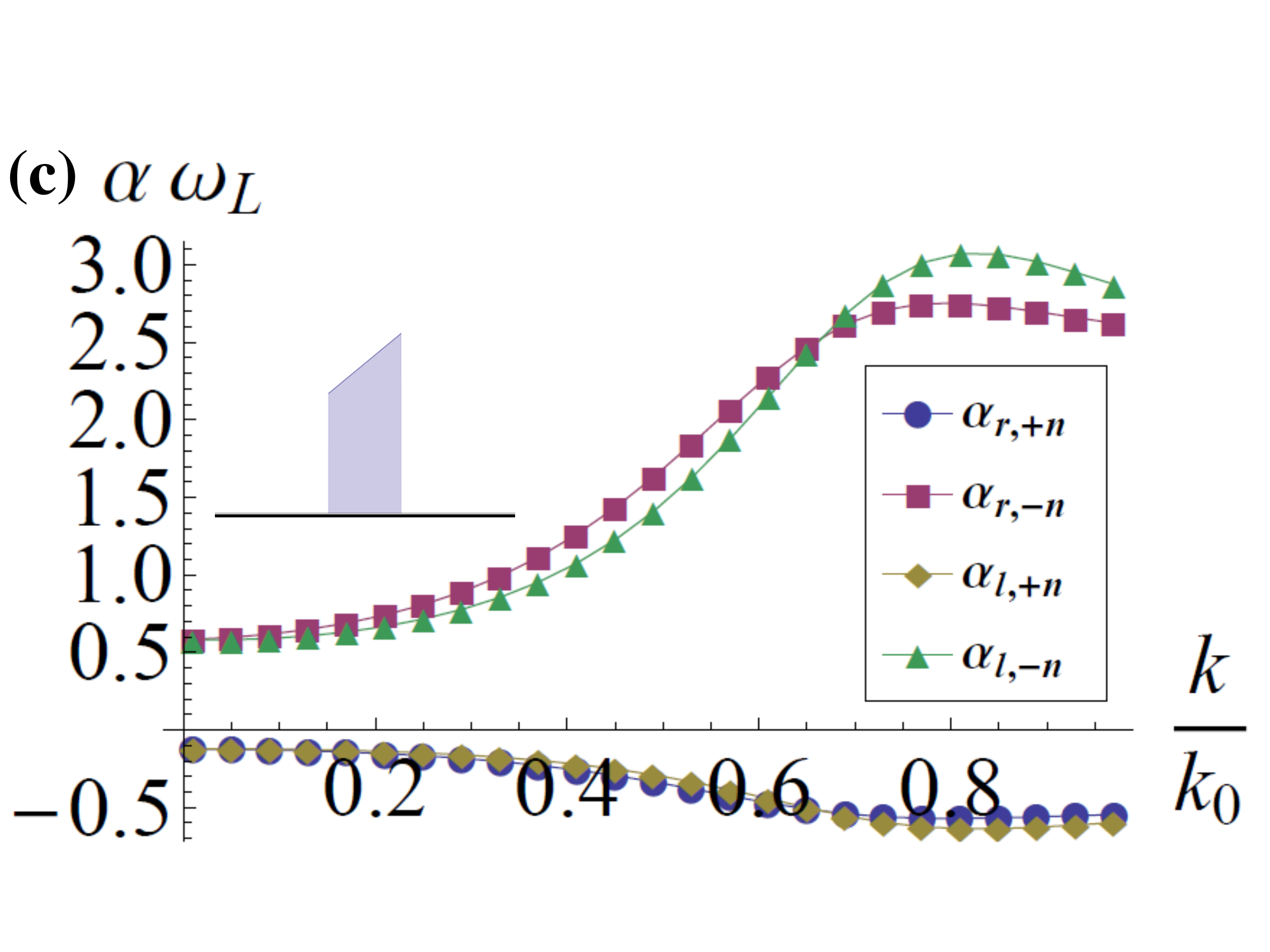}}
\caption{Contextual values times Larmor frequency $\omega_L$ are shown as a function of incident wave vector, for a square (a), a symmetric (b), and a antisymmetric (c) potential barriers. The values of the parameters are $\hbar=1$, $m=1/2$, $dk_0=3\pi$, $a=k_0^2/d^2$, and $\epsilon=0.5k_0^2$. The plots correspond to the spin post-selection in $\theta=\pi/2-\pi/8$ and $\phi=\pi/4$ in Eq. (\ref{arb_spin}).}\label{fig_cv}
\end{center}
\end{figure}

\paragraph{CVs for a square barrier potential.}\label{cvs_for_a_square_barrier}
As an example of the most simple case, we shall consider a square barrier potential, $V(x)=V_0$. In this case, the CVs in Eq. (\ref{cv_elements}) will be simplified by the symmetry of the system, $C_{r,r}=C_{l,l}$, $r^r=r^l$, and $\tau_{yt}=\tau_{yr}$. The transmitted and reflected amplitudes can be decomposed $t=|t|e^{i\phi_t}$ and $r=|r|e^{i\phi_t+i\pi/2}$. The phase difference between them is just $\pi/2$, so the ratio of reflection and transmission amplitudes becomes $r/t=i|r|/|t|$. Moreover, $C_{r,l}$ is a real value for the symmetric barrier. Therefore, we can simplify $\alpha_{r/l}^0$ and $f_{r/l}(n,\delta\tau)$ as
\begin{eqnarray}
\alpha^0\equiv\alpha_{r/l}^0=C_{l,l}-\frac{\tau_{yt}|r|}{\tau_{zt}|t|}C_{r,l},\label{a_00}\\
f_{r}(n,\delta\tau)=-f_l(n,\delta\tau)=\frac{|r|C_{r,l}}{|t|\tau_{zt}}\frac{1}{\textmd{Im}[x_{n}^{(1)}]}.\label{f_r}
\end{eqnarray}
In this special case, we find the following simple relationship between the elements of the dwell time operator and the Larmor times:
\begin{eqnarray}
C_{l,l}(k)&=&\tau_d(k)=\tau_{yt}(k),\nonumber\\
C_{r,l}(k)&=&\frac{m}{\hbar k}\int_{-d/2}^{d/2} dx \phi_{r}(k,x)^*\phi_{l}(k,x)\nonumber\\
&=&\sqrt{\frac{T(k)}{R(k)}}\tau_{zt}(k),\label{dwell_larmor_relations}
\end{eqnarray}
where the first equality can be explained by the fact that the in-plane precession of the spin is same as the average dwell time of the particle in the barrier \cite{Buttiker}.
In contrast to the first relation, the second one Eq. (\ref{dwell_larmor_relations}) works only for the square barrier case, which is not a general relation. This effect further simplifies the equations Eq. (\ref{a_00},\ref{f_r}) to $\alpha^0=0$ and $f_r(n,\delta\tau)=-f_l(n,\delta\tau)=1/\textmd{Im}[x_n^{(1)}]$, which is momentum independent.
Therefore, we see that the CVs have a simple form and only depend on the spin post-selection parameters,
\begin{eqnarray}
\alpha_{r,+n}&=&\alpha_{l,+n}=-\frac{x_{-n}^0}{\omega_L\textmd{Im}[x_n^{(1)}]},\nonumber\\
\alpha_{r,-n}&=&\alpha_{l,-n}=\frac{x_{+n}^0}{\omega_L\textmd{Im}[x_n^{(1)}]}.\label{cvs_squarebarrier}
\end{eqnarray}

In Fig. \ref{fig_cv}(a), we plot the CVs times $\omega_L$ as a function of incident momentum $k$. It shows the momentum independent constant behavior of $\alpha_{r,\pm n}$, whose values are determined by the spin post-selection. Moreover, the weakness of the measurement, controlled by the smallness of $\omega_L$, indicates the measurement is ambiguous: the CVs diverge in order to give the correct average dwell time and tunneling time.

The negative sign of $\alpha_{r,+n}$ in Eq. (\ref{cvs_squarebarrier}) can be understood in the sense that the CVs are determined by the measurement context which is assigned by the experimenter. The negative CVs compensate the ambiguity of the detection to make the average value of the dwell time operator equal to the weighted average of the CVs.

\paragraph{CVs for a non squared symmetric barrier potential.}
For more general symmetric potential barrier, we consider a potential barrier, $V(x)\Theta_B(x)=(V_0+ax^2)\Theta_B(x)$ where $a$ is a real constant. Unlike the square barrier case, $\alpha_{r/l}^0$ in Eq. (\ref{a_00}) is not zero any more and depends also on momentum $k$ because the second relation in Eq. (\ref{dwell_larmor_relations}) is no longer valid. Therefore the CVs in this case are
\begin{eqnarray}
\alpha_{r,+n}&=&\alpha_{l,+n}=\alpha^0-\frac{x_{-n}^0}{\omega_L \textmd{Im}[x_{n}^{(1)}]}\frac{|r|C_{r,l}}{|t|\tau_{zt}},\nonumber\\
\alpha_{r,-n}&=&\alpha_{l,-n}=\alpha^0+\frac{x_{+n}^0}{\omega_L \textmd{Im}[x_{n}^{(1)}]}\frac{|r|C_{r,l}}{|t|\tau_{zt}}.\label{cvs_symmetric}
\end{eqnarray}
The $\omega_L$ dependent deviations of the CVs for $\pm n$ varies for the different spin and momentum.

The CVs are shown in Fig. \ref{fig_cv}(b) for a choice of $a=k_0^2/d^2$, $\theta=\pi/2-\pi/8$ and $\phi=\pi/4$ in Eq. (\ref{arb_spin}). They now have $k$-dependence, and the shape is determined by the momentum dependent factor $(|r|C_{r,l})/(|t|\tau_{zt})$.

\paragraph{CVs for an asymmetric potential barrier.}
Now, we apply our result to a trapezoidal barrier, $V(x)\Theta_B(x)=(V_0+\epsilon(1/2+x/d))\Theta_B(x)$. Since the potential barrier is not symmetric anymore, $\alpha_{r}^0$ and $\alpha_{l}^0$ are not the same and depend both on the spin post-selection $\pm n$ and momentum $k$. The different momentum post-selection leads to all $4$ CVs having different behavior Fig. (\ref{fig_cv}(c)).

\section{Conditioned average of the dwell time operator}\label{conditioned average of the dwell time operator}
The expectation value of the dwell time operator in the initial state $|\psi_{in}\rangle$ incident from the left is independent of the measurement context. The values assigned to $\alpha_{p,m}$ outcomes $p=r,l$ and $m=\pm n$, when averaged with the outcome probabilities $P_{l,p,m}$, will, by construction, give the correct average,
\begin{eqnarray}
\langle\hat{T}_d(k)\rangle&=&\sum_{\substack{p=r,l\\ m=\pm n}} \alpha_{p,m}(k)\textmd{Tr}[\hat{E}_{p,m}(k)\hat{\rho}_{in}]\nonumber\\
&=&C_{l,l}(k)=\tau_d(k), \label{dwell_average}
\end{eqnarray}
guaranteed from Eq. (\ref{cv_decomposition}), where $\hat{\rho}_{in}=|\psi_{in}\rangle\langle\psi_{in}|$ is the initial state from Eq. (\ref{initialstate}). The probability that the particle is prepared initially on the left hand side ($l$) and measured in spin $|m\rangle$ state on the $p$ side of the barrier, is $P_{l,p,m}(k)=\textmd{Tr}[\hat{E}_{p,m}(k)\rho_{in}] $. Although we have considered an incident scattering particle from the left here, we stress that this relation would be equally valid for a particle incident from the right, or indeed, for any coherent combination of left and right initial incoming states.
As an illustration of this check on the derived CVs, we give a detailed derivation of Eq. (\ref{dwell_average}) for the simplest case of a square barrier. Weighting the probabilities $P_{l,p,m}(k)$ by the CVs Eq. (\ref{cvs_squarebarrier}), the average of the dwell time operator is given by
\begin{eqnarray}
\langle\hat{T}_d(k)\rangle&=&\frac{-x_{-n}^0}{\omega_L\textmd{Im}[x_n^{(1)}]}(x_{+n}^0-\omega_L\tau_{yt}\textmd{Im}[x_{n}^{(1)}])\nonumber\\
&&+\frac{x_{+n}^0}{\omega_L\textmd{Im}[x_n^{(1)}]}(x_{-n}^0+\omega_L\tau_{yt}\textmd{Im}[x_{n}^{(1)}])\nonumber\\
&=&\tau_{yt}(k).\label{dwell_average_simple}
\end{eqnarray}
From this simple example, we see why the negative CVs must be there and how the measurement context dependent part of the probabilities and the CVs will cancel out. The probabilities $P_{l,p,\pm n}$ are given by $P_{l,p,\pm n}=(T \textmd{or} R)[x_{\pm n}^0\mp\omega_L\tau_{yt}\textmd{Im}[x_{n}^{(1)}]]$ for $p=r,l$, while the CVs for the same spin post-selection are the same, $\alpha_{r,\pm n}=\alpha_{l,\pm n}$. Consequently, the left/right sum in Eq. (\ref{dwell_average}) gives the $\pm n$ CVs multiplied by $P_{l,l,\pm n}+P_{l,r,\pm n}=x_{\pm n}^0\mp\omega_L\tau_{yt}\textmd{Im}[x_{n}^{(1)}]$. In the spin sum Eq. (\ref{dwell_average_simple}), the negative CV for the $+n$ outcome is responsible for the cancelation of the leading order term. The $\omega_L$ dependence drops out of the remaining term, leaving $\tau_{yt}(x_{+n}^0+x_{-n}^0)=\tau_{yt}$, since $x_{\pm n}^0$ are the leading order probabilities of the spin being found to be $\pm n$.
The average in this measurement context gives the same result as the dwell time $\tau_{d}(k)$ shown in Eq. (\ref{dwell_larmor_relations}).

Now that we see how to construct the normal averages, it is straightforward to post-select the particle.
Our concern is the time taken by the subset of particles that tunnel, and we consider the initially prepared wave packet propagating from the left toward the potential barrier and eventually measured on the transmitted side ($p=r$). This quantity can be formulated by conditioning the average of the CVs.
Since our probability operators already contain the system post-selection, measured on the left or right hand side of the barrier, the conditioned average (for the tunneling case) of the dwell time operator $\hat{T}_d(k)$ is defined simply by choosing the $p=r$ case,
\begin{eqnarray}
_t\langle\hat{T}_d(k)\rangle_{in}&=& \sum_{m=\pm n}\alpha_{r,m}(k)P_{m|l,r}(k)\nonumber\\
&=&\sum_{m=\pm n} \alpha_{r,m}(k) \frac{P_{l,r,m}(k)}{P_{l,r}} ,
\end{eqnarray}
where we use the traditional notation of the conditional probability $P_{m|l,p}=P_{l,p,m}/P_{l,p}$, Bayes' rule. The probability $P_{l,r}=P_{l,r,+n}+P_{l,r,-n}=\textmd{Tr}[\hat{E}_{r,+n}\rho_{in}]
+\textmd{Tr}[\hat{E}_{r,-n}\rho_{in}]$ is the total transmission probability for the initially left coming wave packet. That is, we define the tunneling time as the conditioned average of the CVs. The reflected case is easily calculated analogously to the transmitted one. We note others \cite{Hauge.Stovneng} have stressed that the definition of the tunneling and reflected time should satisfy, $_t\langle\hat{T}_d\rangle T+_r\langle\hat{T}\rangle R=\langle\hat{T}_d\rangle=\tau_{d}$. The CV formalism imposes this condition automatically.

These probabilities can be expressed with the operators
\begin{eqnarray}
_t\langle\hat{T}_d\rangle_{in}=\sum_{\substack{p=r,l\\ m=\pm n}}\alpha_{p,m}\frac{\textmd{Tr}\left[\hat{\mathcal{M}}_{p,m}^{\dag}\hat{F}_{r}\hat{\mathcal{M}}_{p,m} \hat{\rho}_{in}\right]}{\sum_{\substack{p=r,l\\ m=\pm n}}\textmd{Tr}\left[\hat{\mathcal{M}}_{p,m}^{\dag}\hat{F}_{r}\hat{\mathcal{M}}_{p,m} \hat{\rho}_{in}\right]},\label{conditioned_average_k}
\end{eqnarray}
where the $\hat{F}_r$ operator is a projector on the transmitted side that will pick out the $p=r$ term. The detailed calculation up to the first order of $\omega_L$ of the conditional probability for the weak limit lets us understand the conditioned average in terms of known and measurable properties,
\begin{eqnarray}
_t\langle\hat{T}_d\rangle_{in}&=&\sum_{m=\pm n}\alpha_{r,m}\left[ |\langle m|+x\rangle|^2\right.\nonumber\\
&&\qquad\left.+\omega_L\textmd{Re}\left[ (\tau_{zt}-i\tau_{yt})\langle m|+x\rangle\langle -x|m\rangle  \right] \right]\nonumber\\
&=&\alpha_r^0+\frac{\alpha_r^{(1)}}{x_{+n}^0x_{-n}^0}\textmd{Re}\left[ (\tau_{zt}-i\tau_{yt})x_n^{(1)*}\right].\nonumber\\
&=&T_{11}-R\:\textmd{Re}\left[ \delta\tau^*x_n^{(1)}\right] f_r(n,\delta\tau).\label{cond_ave_result}
\end{eqnarray}
The second equality comes from substituting the CVs from Eq. (\ref{cv_rl}), and we see that the result is $\omega_L$ independent but unlike the weak value still depends on the measurement context, as evidenced by the appearance of detector parameters.

To understand Eq. (\ref{cond_ave_result}) physically, let us reformulate Eq. (\ref{conditioned_average_k}) as a sum of two terms,
\begin{eqnarray}
_t\langle\hat{T}_d(k)\rangle_{in}&=&\sum_{\substack{p=a,b\\ m=\pm n}}\alpha_{p,m}\frac{\textmd{Tr}\left[\frac{1}{2}\{\hat{F}_r,\hat{E}_{p,m} \}\hat{\rho}_{in} \right] }{\textmd{Tr}\left[\hat{F}_r\hat{\rho}_{in}\right]}+ D_{n}\nonumber\\
&=&\frac{\textmd{Tr}\left[\frac{1}{2}\{\hat{F}_r,\hat{T}_d(k) \}\hat{\rho}_{in} \right]}{\textmd{Tr}\left[\hat{F}_r\hat{\rho}_{in}\right]}+D_{n},\label{weak_distur}
\end{eqnarray}
where first term is the (context independent) weak limit of the conditioned average and $D_{n}$ is the context dependent disturbance
\begin{flalign}
&&D_{n}=\sum_{\substack{p=a,b\\ m=\pm n}}\frac{\alpha_{p,m}}{\textmd{Tr}\left[\hat{F}_r\hat{\rho}_{in}\right]}\textmd{Re}\left( \textmd{Tr}\left[ [\hat{\mathcal{M}}_{p,m}^{0\dag},\hat{F}_r]\hat{\mathcal{M}}_{p,m}^0\hat{\rho}_{in}\right.\right.\nonumber\\
&&\quad\left.\left.+\omega_L \{ [\hat{\mathcal{M}}_{p,m}^{0\dag},\hat{F}_r]\hat{\mathcal{M}}_{p,m}^{(1)}\hat{\rho}_{in}
+\hat{\mathcal{M}}_{p,m}^{0\dag}[\hat{F}_r,\hat{\mathcal{M}}_{p,m}^{(1)}]\hat{\rho}_{in} \} \right] \right),\label{disturbance}
\end{flalign}
where we defined the operators
\begin{eqnarray}
\hat{\mathcal{M}}_{p,m}^{0}&\equiv&\hat{\Pi}_{p}\hat{M}_{m}^0,\\
\hat{\mathcal{M}}_{p,m}^{(1)}&\equiv&\hat{\Pi}_{p}\hat{M}_{m}^{(1)},
\end{eqnarray}
from Eq. (\ref{measurement_operator},\ref{projection_operator}).
For a pure initial state $\hat{\rho}_{in}$ and the post-selection projector $\hat{F}_r$, the first term of Eq. (\ref{weak_distur}) simplifies to the real part of the weak value, $\textmd{Re}\left[ T_d^w(k) \right]$, which is a general property of the CV formalism in \cite{dressel0}
\begin{eqnarray}
\textmd{Re}\left[ T_d^w(k) \right]&=&\frac{\textmd{Tr}\left[\frac{1}{2}\{\hat{F}_r,\hat{T}_d(k) \}\hat{\rho}_{in} \right]}{\textmd{Tr}\left[\hat{F}_r\hat{\rho}_{in}\right]}\\
&=&\lim_{\tau\rightarrow\infty}\textmd{Re}\left[ \frac{\langle k(\tau)|\hat{T}_d(k)|k(-\tau)\rangle}{\langle k(\tau)|k(-\tau)\rangle} \right].\nonumber
\end{eqnarray}
The disturbance term $D_{n}$, on the other hand, comes from the non-commutativity of the measurement operator $\hat{\mathcal{M}}_{p,m}$ and the post-selection projector $\hat{F}_r$. The joint probability in Eq. (\ref{conditioned_average_k}) contains information not only about the measurement and the initial state, but also about the post-selection and the disturbance to the initial state due to the measurement. Therefore, due to the freedom of post-selection basis, the disturbance $D_{n}$ could be maximized or minimized as the experimenter desires.

\begin{figure}
\begin{center}
\subfigure{\includegraphics[width=7cm]{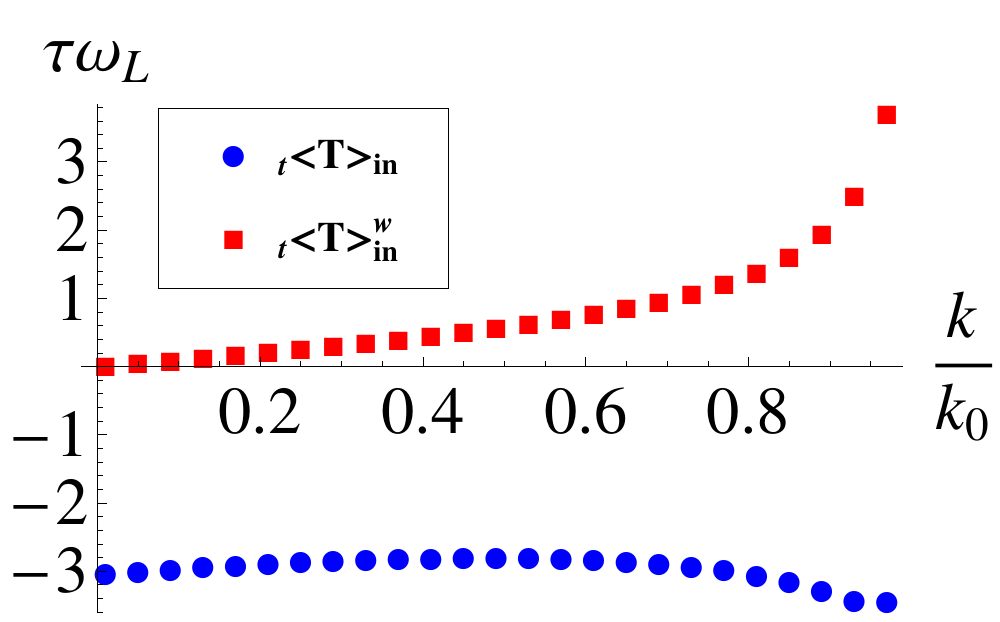}}
\subfigure{\includegraphics[width=7cm]{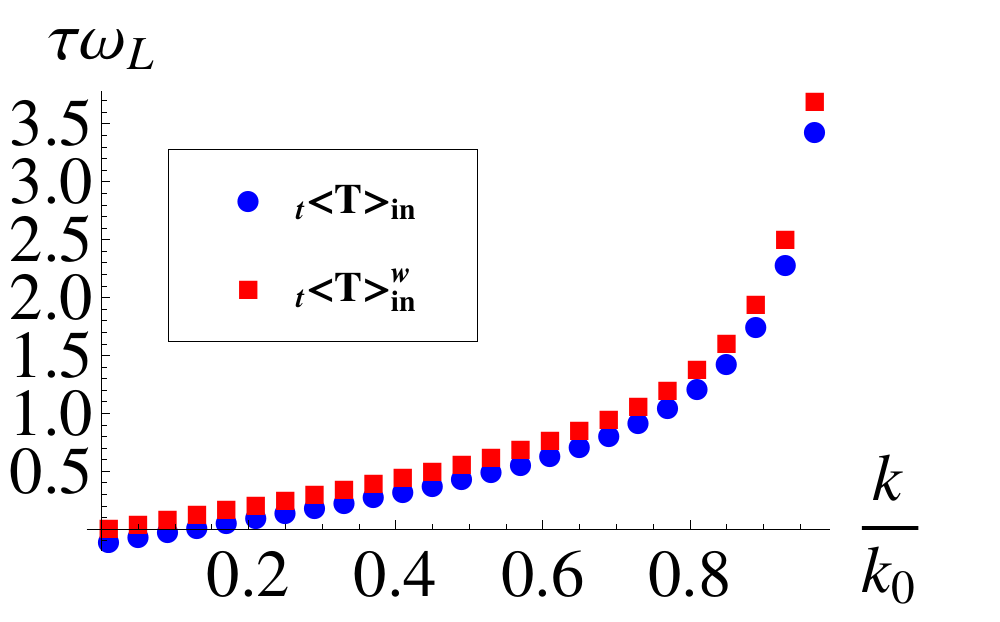}}
\subfigure{\includegraphics[width=7cm]{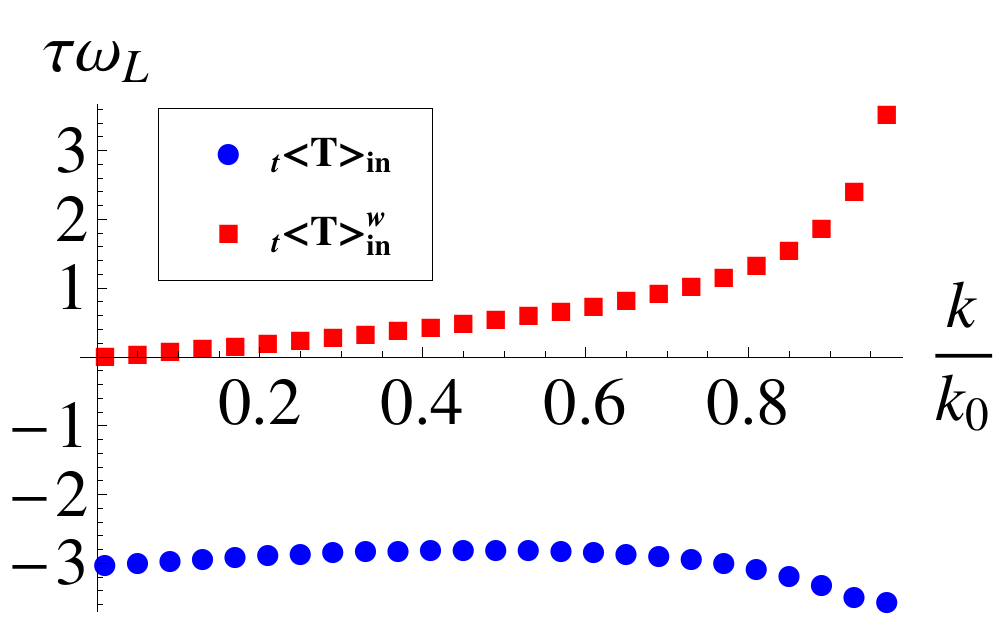}}
\caption{The real part of the weak value and the conditioned average are compared for different shapes of the symmetric potential barrier. (a) and (b) are for the square barrier, corresponding to the spin post-selection, $(\theta,\phi)=(\pi/2-\pi/8,\pi/4)$ and $(\theta,\phi)=(\pi/2-\pi/200,\pi/4)$, respectively. (c) is for the non square symmetric potential as in Fig. (\ref{fig_cv}(b)) when $(\theta,\phi)=(\pi/2-\pi/8,\pi/4)$. The values of the parameters are $\hbar=1$, $m=1/2$, $dk_0=3\pi$, and $a=k_0^2/d^2$. }\label{fig_times}
\end{center}
\end{figure}

As a special case, when we consider the symmetric potential barrier, the phases of the reflected amplitude from the right and left hand side of the barrier are the same; some of the Larmor times become equal $\tau_{yr}^r=\tau_{yr}^l=\tau_{yt}$ and the difference of the complex times $\delta\tau=\tau_{t}-\tau_{r}^l=\tau_{zt}-\tau_{zr}=\tau_{zt}/R$ is real. Therefore, when we use these properties together with the relations Eq. (\ref{f_r}), the conditioned average Eq. (\ref{cond_ave_result}) takes a simple form,
\begin{eqnarray}
_t\langle\hat{T}_d\rangle_{in}=\textmd{Re}\left[T_d^w(k)\right]\label{ca_symmetric}
-\frac{|r|}{|t|}C_{r,l}\frac{\textmd{Re}[x_n^{(1)}]}{\textmd{Im}[x_n^{(1)}]}.
\end{eqnarray}
As we expected, the first term of the right hand side is the same as the real part of the weak value Eq. (\ref{weakvalue_formalism}), $\textmd{Re}\left[T_d^w(k)\right]=C_{l,l}=\tau_{d}$ because $\textmd{Re}[(r/t)C_{r,l}]=0$. The second term in Eq. (\ref{ca_symmetric}) is the detector parameter dependent disturbance which has a simple form in terms of the off diagonal element of the dwell time operator $C_{r,l}$ weighted by the ratio of $|r|/|t|$ and the post-selection amplitudes of the detector $\textmd{Re}[x_n^{(1)}]/\textmd{Im}[x_n^{(1)}]$. The conditioned average is bounded by the CVs in Eq. (\ref{cvs_symmetric}). Since the range of the CVs is larger than the eigenvalues of $\hat{T}_d$ \cite{Muga2} due to the amplification from the measurement ambiguity, the conditioned averages can in principle lie anywhere within the CVs range. In the weak coupling limit, $\omega_L\rightarrow 0$, the range of the CVs in Fig. (\ref{fig_cv}) diverge, and the experimental result could be obtained even in the negative time region depending on the choice of the detector parameters, shown in Fig. (\ref{fig_times}). This simply corresponds to the conditional probabilities enhancing the negative CV over the positive one.

In the case of a square barrier potential, we can also use the relations Eq. (\ref{dwell_larmor_relations}), so Eq. (\ref{ca_symmetric}) simplifies further to
\begin{eqnarray}
_t\langle\hat{T}_d\rangle_{in}=\tau_{yt}-\tau_{zt}\frac{\textmd{Re}[x_n^{(1)}]}{\textmd{Im}[x_n^{(1)}]}.\label{cond_ave_symm}
\end{eqnarray}
In this special case, the weak value becomes $\tau_{yt}$ and the disturbance correction has a simple form in terms of $\tau_{zt}$ weighted by the spin overlap parameters. Note that $\tau_{yt}$ is a shift of the detector's pointer and the $\tau_{zt}$ is a measure of the back action on the particle due to the measurement interaction \cite{Steinberg}.
The measurement operator $\hat{M}_{p,m}$ disturbs the initial system state or the system post-selection (in Eq. (\ref{disturbance}) we show the disturbance to the post-selection), but the choice of the tuning parameter $x_n^{(1)}=\langle+x|+n\rangle\langle+n|-x\rangle$ can control the disturbance correction to the tunneling time \cite{dressel2}. Therefore, the proper choice of the tuning parameter can minimize the disturbance correction.
When $\hat{n}$ approaches the $x-y$ plane, $x_n^{(1)}\rightarrow 0$, we can make the disturbance part negligible compared to $\tau_{yt}$. This causes the system to monitor only the precession of the Larmor clock and reduces the disturbance. On the other hand, if we take $\hat{n}$ approaching the $y-z$ plane, the correction diverges. In Fig. \ref{fig_times}(a),(b), we compare the weak value $\textmd{Re}\left[T_d^w(k)\right]$ with $_t\langle\hat{T}_d\rangle_{in}$ for the square barrier. For the spin post-selection $\theta=\pi/2-\pi/8$ and $\phi=\pi/4$, Fig. \ref{fig_times}(a) shows the negative values of the conditioned average which comes from the large pre-factor $(\theta,\phi)$ in the post-selection. In the other limit, when the spin post-selection approaches the $x-y$ plane, $\theta=\pi/2-\pi/200$ and $\phi=\pi/4$, Fig. \ref{fig_times}(b) shows that the disturbance part is negligible and the conditioned average is almost same as the weak value.

\subsection{Second moment of an observable}\label{Second_moment}
The higher moments of the dwell time operator can also be obtained by this measurement process. For instance, the second moment of the operator can be found from the CV formalism. It is well known that the square of the weak value is not the weak value of the square of the operator. From Eq. (\ref{cv_decomposition}), the square of the dwell time operator can be written
\begin{eqnarray}
\hat{T}_d^2(k)=\sum_{\substack{p,p'=r,l\\ m,m'=\pm n}}\alpha_{p,m}(k)\alpha_{p',m'}(k)\hat{E}_{p,m}(k)\hat{E}_{p',m'}(k),
\end{eqnarray}
and the second moment is determined by the average of the two probability operators with the initial state, $\textmd{Tr}[\hat{E}_{p,m}(k)\hat{E}_{p',m'}(k)\hat{\rho}_{in}]$. However, the measurement operators do not commute with each other in general, and $\textmd{Tr}[\hat{E}_{p,m}(k)\hat{E}_{p',m'}(k)\hat{\rho}_{in}]$ is not always a measurable probability.
The probability of two consecutive measurements $\textmd{Tr}[\hat{M}_{p,m}^{(1)}\hat{M}_{p',m'}^{(2)}\hat{F}_r\hat{M}_{p',m'}^{(2)\dag}\hat{M}_{p,m}^{(1)\dag}\hat{\rho}_{in}]$ can be a different quantity.
Therefore, a sequence of two consecutive measurements does not  generally construct the second moment of the dwell time operator.
Instead, another strategy can be employed by changing the CVs to $\beta_{p,m}(k)$ to define a new operator that correspond to powers of the original observable,
\begin{eqnarray}
\hat{T}^2_d(k)=\sum_{\substack{p=r,l\\ m=\pm n}}\beta_{p,m}(k)\hat{E}_{p,m}(k).
\end{eqnarray}
The second moment of the dwell time operator, then, can be measured using the same experimental setup without sequential measurements. The matrix elements of the squared operator are
\begin{eqnarray}
\hat{T}^2_d(k)&=&\left(
                 \begin{array}{cc}
                   \bar{C}_{1,1} & \bar{C}_{1,2} \\
                   \bar{C}_{2,1} & \bar{C}_{2,2} \\
                 \end{array}
               \right)\\
               &=&\left(
                  \begin{array}{cc}
                    C_{l,l}^2+|C_{r,l}|^2 & C_{l,r}(C_{l,l}+C_{r,r}) \\
                    C_{r,l}(C_{l,l}+C_{r,r}) & C_{r,r}^2+|C_{r,l}|^2 \\
                  \end{array}
                \right).\nonumber
\end{eqnarray}
Since $\hat{T}_d(k)$ and $\hat{T}^2_d(k)$ are diagonal in the same basis, a set of CVs, $\{\beta_{p,m}(k)\}$ can be easily obtained by putting the matrix elements $\bar{C}_{\bar{i},\bar{j}}$ instead of $C_{i,j}$ into Eq. (\ref{matrix}), where $(\bar{i},\bar{j})\in(1,2)$ and $(i,j)\in(l,r)$. As an example, the CVs of $\hat{T}_d^2(k)$ for the symmetric potential barrier are,
\begin{eqnarray*}
\beta_{r,+n}&=&\bar{C}_{1,1}-\frac{\tau_{yt}|r|}{\tau_{zt}|t|}\bar{C}_{2,1}
-\frac{x_{-n}^0}{\omega_L\textmd{Im}[x_n^{(1)}]}\frac{|r|\bar{C}_{2,1}}{|t|\tau_{zt}},\\
\beta_{r,-n}&=&\bar{C}_{1,1}-\frac{\tau_{yt}|r|}{\tau_{zt}|t|}\bar{C}_{2,1}
+\frac{x_{+n}^0}{\omega_L\textmd{Im}[x_n^{(1)}]}\frac{|r|\bar{C}_{2,1}}{|t|\tau_{zt}},
\end{eqnarray*}
and as we expected $\beta_{r,\pm n}=\beta_{l,\pm n}$. These are not simple powers of $\alpha_{p,m}$ as seen from Eq. (\ref{cvs_symmetric}).
A measurement of $\langle\hat{T}_d(k)\rangle$ and $\langle\hat{T}_d^2(k)\rangle$ permits us to find the uncertainty of the averaged dwell time expressed as
\begin{eqnarray}
\Delta\hat{T}_d(k)=\sqrt{\langle\hat{T}_d^2(k)\rangle-\langle\hat{T}_d(k)\rangle^2}.
\end{eqnarray}
This procedure easily extends to the $n^{th}$ moment using only data from the spin measurements.


\section{Comparison with Steinberg's approach}\label{Comparison_with_another_idea}
Since our approach is similar to Steinberg's paper \cite{Steinberg}, we want to compare his result with ours. Let us quickly review his idea \cite{Steinberg,Steinberg2}.
For a symmetric barrier and symmetric initial conditions, $\psi_t$, the state of a transmitted particle is simply obtained by a parity flip combined with time reversal $\psi_t(x,\tau)=\psi_i(-x,-\tau)^*$, where $\psi_i(x,\tau)$ is the initial state in which the particle is prepared. In more practical terms, he defined $\psi_t(x,\tau)=t^*\psi_i(x,\tau)+r^*\psi_i(-x,\tau)$. Strictly speaking for a specified $k$, $\psi_i\rightarrow \phi_l(x,\tau)$, where
\begin{eqnarray*}
\phi_l(x,\tau)&=&\frac{1}{\sqrt{2\pi}}\left\{
                             \begin{array}{ll}
                             (e^{ikx}+r e^{-ikx}) e^{-i\omega \tau}, & x < -d/2 \\
                             t e^{ikx} e^{-i\omega\tau}, & x > d/2
                             \end{array}
                             \right.,\\
\phi_r(x,\tau)&=&\frac{1}{\sqrt{2\pi}}\left\{
                             \begin{array}{ll}
                             t e^{-ikx} e^{-i\omega\tau}, & x < -d/2 \\
                             (e^{-ikx}+r e^{ikx})e^{-i\omega\tau}, & x > d/2
                             \end{array}
                             \right.,
\end{eqnarray*}
are the scattering states of the incoming stationary scattering case and we have defined $\phi_{l/r}(x,\tau)=\langle x|k_{l/r}\rangle$. The transmitted state is therefore,
\begin{eqnarray}
\psi_t(x,t)&=& t^*\phi_l(x,\tau)+r^*\phi_r(x,\tau)\nonumber\\
&=&\frac{1}{\sqrt{2\pi}}\left\{
          \begin{array}{ll}
          t^*e^{ikx} e^{-i\omega\tau} , & x < -d/2 \\
          (e^{ikx}+r^* e^{-ikx})e^{-i\omega\tau}, & x > d/2
          \end{array}
          \right.,\nonumber\\
&=&\psi_i(-x,-t)^*,
\end{eqnarray}
where we use the symmetry $\phi_l(-x,\tau)=\phi_r(x,\tau)$. Now the weak value of the projector onto the barrier region $\hat{\Theta}_B=\int_{-d/2}^{d/2}|x\rangle\langle x|$ is used to define Steinberg's transmission time,
\begin{eqnarray}
\tau_{t,s}&\equiv& \frac{m}{\hbar k}\frac{\langle\psi_t|\hat{\Theta}_B|\psi_i\rangle}{\langle\psi_t|\psi_i\rangle}=\frac{m}{\hbar k}\frac{\int_{-d/2}^{d/2}dx\:\tilde{\phi}_r\tilde{\phi}_l}{t}\nonumber\\
&=& \tau_d-i\tau_{i,s},\label{stein_time}
\end{eqnarray}
where $\hbar k/m=j_{in}$ is the incoming current density, and we easily see $\langle\psi_t|\psi_i\rangle=t$. The real part of the transmission time is $\tau_d$ which is the dwell time defined in Eq. (\ref{dwelltime(k)}), and we define $\tau_{i,s}$ as the imaginary part of it.

To compare our result with Steinberg's, we do the same pre- and post-selection with the system operator $\hat{\Theta}_B=\int_{-d/2}^{d/2}|x\rangle\langle x|$ but with the unitary evolution in the region that we are interested in,
\begin{eqnarray*}
\hat{T}_D &=& \int_{-\infty}^{\infty}d\tau e^{i\hat{H}\tau/\hbar}\int_{-d/2}^{d/2} dx|x\rangle \langle x| e^{-i\hat{H}\tau/\hbar}\\
&=&\frac{m}{\hbar k} \int_{0}^{\infty} dk \left[ \sum_{i,i'=r,l} |k_i\rangle \langle k_{i}|\hat{\Theta}_B|k_{i'}\rangle  \langle k_i'| \right],\\
\hat{T}_d(k)&=& \frac{m}{\hbar k}\left[ |k_l\rangle  \langle k_{l}|\hat{\Theta}_B|k_{l}\rangle \langle k_l|+|k_r\rangle\langle k_{r}|\hat{\Theta}_B|k_{l}\rangle\langle k_l|\right.\nonumber\\
&&\left.+|k_l\rangle\langle k_{l}|\hat{\Theta}_B|k_{r}\rangle\langle k_r|+|k_r\rangle\langle k_{r}|\hat{\Theta}_B|k_{r}\rangle\langle k_r| \right].
\end{eqnarray*}
Therefore, the weak value of the operator $\hat{T}_d$ is
\begin{eqnarray*}
\frac{\langle\psi_t|\hat{T}_d|\psi_i\rangle}{\langle\psi_t|\psi_i\rangle}
&=&\frac{m}{\hbar k}\left[  \int_{-d/2}^{d/2}dx\:\tilde{\phi}_l^*\tilde{\phi}_l+\frac{r}{t}\int_{-d/2}^{d/2}dx\:\tilde{\phi}_r^*\tilde{\phi}_l  \right]\\
&=& \tau_d+\frac{r}{t} C_{r,l},
\end{eqnarray*}
where we use the dwell time definition $\tau_d=C_{l,l}$ and find the same result as our weak value Eq. (\ref{weakvalue_formalism}).

For the symmetric barrier, the weak value simplifies to
\begin{eqnarray}
\frac{\langle\psi_t|\hat{T}_d|\psi_i\rangle}{\langle\psi_t|\psi_i\rangle}=\tau_d+i \frac{|r|}{|t|} C_{r,l}.
\end{eqnarray}
by applying the properties of the symmetric barrier.
The conditioned average of the dwell time is given by the real part of the weak value, and the real part of our result is the same as Steinberg's result Eq. (\ref{stein_time}). However the imaginary parts are not equal because we use the dwell time operator and not a scaled projector. Since the imaginary part is related to the back action of the detector \cite{Steinberg,dressel2}, these are not expected to be the same.

\section{Conclusion}\label{Conclusion}
We have put forward a principled approach to making indirect measurements of dwell and tunneling times. The starting point for our work is defining the dwell time as a self-adjoint time operator which is the observable for the time spent in a spatial region. An operational approach is then taken, where this operator is indirectly measured with the help of measurements made on an auxiliary spin degree of freedom, which weakly interacts with a magnetic field in the region of interest. We give a prescription for finding the expectation of the operator by assigning CVs to all the outcomes of the spin measurements on the reflected and transmitted side of the barrier, focusing on the case of particles at a fixed energy for simplicity. These CVs encode the physics of the scattering process. When the CVs are averaged with the outcome frequencies of their events, they produce the expectation of the dwell time operator, regardless of the initial state. We find their form explicitly for a general one dimensional barrier of finite extent. Interestingly, in order to have well defined CVs, the post-selection angle for the spin must be such that the probability of detector outcomes depends on both changes in magnitude and phase of the scattering amplitudes in order to have a full reconstruction of the dwell time operator.

To define the tunneling time, that is, the dwell time of the particles which are post-selected on the transmitted side of the barrier, we average the same CVs, but now with the conditional probabilities of the detector results, where the conditioning is on the successful tunneling events. With this definition, we recover the weak value of the full dwell time operator as the tunneling time, plus a detector-dependent disturbance term. This disturbance depends on the choice of post-selection basis, and we find it can be made negligibly small by choosing to measure the spin in a basis nearly orthogonal to the magnetic field direction. In the simplest case of a square barrier, the tunneling time is simply the in-plane portion of the Larmor time. The strength of this line of research into tunneling time is that the results are immediately applicable to experiments, once the CVs are calibrated and assigned to the experimental outcomes.

\section*{Acknowledgment}
We acknowledge helpful discussions with Justin Dressel and support from the National Science Foundation under Grant No. DMR-$0844899$


\appendix

\section{The dwell time operator}
By definition, the dwell time operator is
\begin{eqnarray}
\hat{T}_D&=&\int_{-\infty}^{\infty}dt e^{i\hat{H}t/\hbar}\int_{-d/2}^{d/2}dx|x\rangle\langle x|e^{-i\hat{H}t/\hbar} \label{dwell_op_def}\\
&=&\int_{-\infty}^{\infty}dt e^{i\hat{H}t/\hbar}\:\hat{\chi}_d\: e^{-i\hat{H}t/\hbar}.\nonumber
\end{eqnarray}
For simplicity, we defined the projector onto the region $[-d/2,d/2]$,
\begin{eqnarray}
\hat{\chi}_{d}=\int_{-d/2}^{d/2}dx |x\rangle\langle x|.
\end{eqnarray}
Since the time integration of the dwell time operator goes from negative infinity to infinity, the commutation of the time operator and the Hamiltonian $[\hat{T}_D,\hat{H}]=0$ can be clearly shown \cite{Muga2}
\begin{eqnarray}
\hat{T}_De^{-i\hat{H}t/\hbar}&=&\int_{-\infty}^{\infty}d\tau e^{i\hat{H}\tau/\hbar}\:\hat{\chi}_d\: e^{-i\hat{H}(\tau+t)/\hbar}\nonumber\\
&=&\int_{-\infty}^{\infty}d\tau e^{i\hat{H}(\tau-t)/\hbar}\:\hat{\chi}_d\: e^{-i\hat{H}\tau/\hbar}\nonumber\\
&=&e^{-i\hat{H}t/\hbar}\:\hat{T}_D.\label{commutation}
\end{eqnarray}
Therefore, the Eq. (\ref{dwell_op_def}) can be written in the scattering basis of the Hamiltonian,
\begin{eqnarray}
\hat{T}_D=\int_{0}^{\infty} dk \sum_{i,i'=r,l}C_{i,i'}(k)|k_i\rangle\langle k_{i'}|, \label{dwell_op}
\end{eqnarray}
which is the momentum resolved dwell time operator.
If we prepare the state of the system $|\Psi\rangle=\alpha|k_l\rangle+\beta|k_r\rangle$, the average of the momentum resolved dwell time operator in this state is
\begin{eqnarray*}
\langle\Psi|\hat{T}_d (k)|\Psi\rangle=|\alpha|^2C_{l,l}+|\beta|^2C_{r,r}+\alpha\beta^*C_{r,l}+\alpha^*\beta C_{l,r}.
\end{eqnarray*}
The diagonal element $C_{l,l}$ ($C_{r,r}$) is the dwell time for the initially left (right) moving state. The off diagonal elements $C_{r,l}$ appear for an initially prepared left moving scattering state and has a transition to the right moving scattering state. Therefore, this expression shows the explicit meaning of the averaged dwell time for a coherent superposition of left and right moving states.

We can easily find the eigenvalues and the corresponding (un-normalized) eigenstates $\hat{T}_d(k)|\lambda_{\pm}\rangle=\lambda_{\pm}|\lambda_{\pm}\rangle$,
\begin{widetext}
\begin{eqnarray*}
\lambda_{\pm}(k)&=&\frac{1}{2}\left[C_{r,r}(k)+C_{l,l}(k)\pm\sqrt{(C_{r,r}(k)-C_{l,l}(k))^2+4 |C_{r,l}(k)|^2} \right],\label{eigenvalues}\\
|\lambda_{+}(k)\rangle&=&\left[ \left(C_{l,l}(k)-C_{r,r}(k)+\sqrt{(C_{l,l}(k)-C_{r,r}(k))^2+4 |C_{r,l}(k)|^2}\right)|k_l\rangle+ 2C_{r,l}(k)|k_{r}\rangle \right],\nonumber \\
|\lambda_{-}(k)\rangle&=&\left[ 2C_{r,l}^*(k)|k_{l}\rangle -\left(C_{l,l}(k)-C_{r,r}(k)+\sqrt{(C_{l,l}(k)-C_{r,r}(k))^2+4 |C_{r,l}(k)|^2}\right)|k_r\rangle \right]. \label{eigenstates}
\end{eqnarray*}
\end{widetext}
The eigenvalues for a box potential barrier are worked in  Ref. \onlinecite{Muga2}. When we consider the higher moments of the dwell time operator, $\langle \hat{T}_D^{n}\rangle$ it is easy to compute in the eigensystem.

\section{Orthonormality of the scattering states}
Consider the left and right incoming Hamiltonian eigenstates, $k>0$,
\begin{eqnarray}
\phi_l(x,k)=\frac{1}{\sqrt{2\pi}}\left\{
                             \begin{array}{ll}
                             e^{ikx}+r^l e^{-ikx}, & x < -a \\
                             t e^{ikx}, & x > a
                             \end{array}
                             \right.,\nonumber\\
\phi_r(x,k)=\frac{1}{\sqrt{2\pi}}\left\{
                             \begin{array}{ll}
                             t e^{-ikx} , & x < -a \\
                             e^{-ikx}+r^r e^{ikx}, & x > a
                             \end{array}
                             \right.,\label{eigenbasis}
\end{eqnarray}
where the transmission and reflection amplitudes are $k$ dependent. We want to show that these are orthonormal.

\subsection{Orthonormality of the scattering states.}\label{normalization}
To prove the left scattering states are orthonormal, let us start by calculating the inner product
\begin{eqnarray*}
\langle \phi_l|\phi_l'\rangle&=&\int_{-\infty}^{\infty}dx\:\phi_l(x,k)^*\phi_l(x,k')\\
&=&\frac{1}{2\pi}\int_{-\infty}^{-a}dx(e^{-ikx}+r^{l^*}e^{ikx})(e^{ik'x}+r^{l'}e^{-ik'x})\\
&&+\frac{1}{2\pi}\int_a^{\infty}dx\: t^*t'e^{-i(k-k')x}\\
&&+\int_{-a}^a dx\: \phi_l^*(x,k)\phi_l(x,k')\\
&=& g_1+g_2+g_3+g_4+\int_{-a}^a dx\: \phi_l^*(x,k)\phi_l(x,k'),
\end{eqnarray*}
where we consider $k'$ nearby $k$, $k'=k+\epsilon$ for $\epsilon\rightarrow 0$, and we substitute $\lambda\rightarrow \infty$ instead of $\infty$ in the integration limits. Then the functions, $g_1, g_2, g_3$, and $g_4$, are defined as
\begin{eqnarray*}
g_1&\equiv&\lim_{\epsilon\rightarrow 0}\lim_{\lambda\rightarrow\infty} \frac{e^{-i\epsilon\lambda}-e^{-i\epsilon a}}{-2\pi i\epsilon},\\
g_2&\equiv&\lim_{\epsilon\rightarrow 0}\lim_{\lambda\rightarrow\infty}r^{l^*}\frac{e^{-i(2k+\epsilon)\lambda}-e^{-i(2k+\epsilon)a}}{-2\pi i(2k+\epsilon)},\\
g_3&\equiv&\lim_{\epsilon\rightarrow 0}\lim_{\lambda\rightarrow\infty}r^{l'} \frac{e^{i(2k+\epsilon)\lambda}-e^{i(2k+\epsilon)a}}{2\pi i(2k+\epsilon)},\\
g_4&\equiv&\lim_{\epsilon\rightarrow 0}\lim_{\lambda\rightarrow\infty}(r^{l^*}r^{l'}+t^*t')\frac{e^{i\epsilon\lambda}-e^{i\epsilon a}}{2\pi i\epsilon}.
\end{eqnarray*}
For small $\epsilon$, the amplitudes $t'$ and $r^{l'}$ can be expanded around $k$ to first order in $\epsilon$
\begin{eqnarray*}
t'&\simeq& t+\epsilon\:\partial_k t,\\
r^{l'}&\simeq& r^l+\epsilon\:\partial_k r^l.
\end{eqnarray*}
The sum $g_2+g_3$ becomes
\begin{eqnarray*}
g_2+g_3=\frac{1}{4\pi ik}(r^{l^*}e^{-2ika}-r^le^{2ika}),
\end{eqnarray*}
in this limit. Since the amplitudes satisfy $|t|^2+|r|^2=1$, the sum $g_1+g_4$ is given by
\begin{eqnarray*}
g_1+g_4&=&\frac{i}{2\pi}(r^{l*}\partial_k r^l+t^*\partial_k r)\\
&&+\lim_{\epsilon\rightarrow 0}\lim_{\lambda\rightarrow\infty}(\frac{e^{i\epsilon\lambda}-\epsilon^{i\epsilon a}}{2\pi i\epsilon}+c.c).
\end{eqnarray*}
Now, the first part of the second term of the above equation can be rewritten as
\begin{eqnarray*}
&&\lim_{\epsilon\rightarrow 0}\lim_{\lambda\rightarrow\infty}\frac{e^{i\epsilon\lambda}-\epsilon^{i\epsilon a}}{2\pi i\epsilon}\\
&&=\lim_{\epsilon\rightarrow 0}\lim_{\lambda\rightarrow\infty}e^{\frac{i}{2}(\epsilon\lambda+\epsilon a)}\frac{e^{\frac{i}{2}(\epsilon\lambda-\epsilon a)}-e^{-\frac{i}{2}(\epsilon\lambda-\epsilon a)}}{2\pi i\epsilon}\\
&&=\lim_{\epsilon\rightarrow 0}\lim_{\lambda\rightarrow\infty}\frac{\sin(\frac{\epsilon}{2}(\lambda-a))}{\pi\epsilon}\\
&&=\lim_{\epsilon\rightarrow 0}\lim_{\lambda\rightarrow\infty}\frac{\lambda-a}{2\pi}\textmd{sinc}(\frac{\epsilon}{2\pi}(\lambda-a))= \lim_{\epsilon\rightarrow 0}\frac{1}{2}\delta(\epsilon).
\end{eqnarray*}
Therefore, the inner product is
\begin{eqnarray}
&&\langle \phi_l|\phi_l'\rangle\nonumber\\
&&=-\frac{1}{4\pi ik}(r^{l^*}e^{-2ika}+r^le^{2ika})+\frac{i}{2\pi}(r^{l*}\partial_k r^l+t^*\partial_k r) \nonumber\\
&&+\delta(k-k')+\int_{-a}^a dx\: \phi_l^*(x,k)\phi_l(x,k').
\end{eqnarray}

To finish the proof that $\langle\phi_l|\phi_l'\rangle=\delta(k-k')$, let us consider the Schr\"{o}dinger equation,
\begin{eqnarray}
H\phi&=&(T+V)\phi=E\phi,\nonumber\\
(H-E)\phi&=&(T+V-E)\phi=0,\label{Hamiltonian_Equation}
\end{eqnarray}
here $H$ is Hamiltonian, $T$ is kinetic energy, and $V$ is potential energy.
The first derivative with respect to $E$ is \cite{Smith},
\begin{eqnarray*}
\partial_E\left[(H-E)\phi \right]=(H-E)\partial_E\phi-\phi=0.
\end{eqnarray*}
For a local potential and for the same coordinates in the two eigenfunctions, this equation gives
\begin{eqnarray*}
&&\phi_l^*T\left[\partial_E\phi_l\right]-\left[\partial_E \phi_l\right]\left[T\phi_l\right]^*\\
&&=\phi_l^*\left[ \phi_l-(V-E)\partial_E\phi_l\right]-\left[ \partial_E\phi_l\right]\left[T\phi_l\right]^*\\
&&=\phi_l^*\phi_l-\phi_l^*\left[V-E\right]\left[\partial_E\phi_l\right]+\left[\partial_E\phi_l\right]\left[V-E\right]\phi_l^*\\
&&=\phi_l^*\phi_l.
\end{eqnarray*}
This equation is true only when $x_1=x_2$ and $k_1=k_2$ for $\phi_l^*(x_1,k_1)\phi_l(x_2,k_2)$, since $V(x_1)\neq V(x_2)$ in general. Now we find
\begin{eqnarray*}
\phi_l^*\phi_l&=&\phi_l^*T(\partial_E\phi_l)-(\partial_E \phi_l)T\phi_l^*\\
&=&\phi_l^*\left(-\frac{\hbar^2}{2m}\partial_x^2\right)\partial_E\phi_l-(\partial_E\phi_l)\left(-\frac{\hbar^2}{2m}\partial_x^2\right)\phi_l^*\\
&=&-\frac{\hbar^2}{2m}\left[ \phi_l^*\partial_x^2(\partial_E\phi_l)-(\partial_E\phi_l)(\partial_x^2\phi_l^*) \right]\\
&=&-\frac{\hbar^2}{2m}\partial_x\left[ \phi_l^*(\partial_x\partial_E\phi_l)-(\partial_E\phi_l)(\partial_x\phi_l^*) \right].
\end{eqnarray*}
Integration from $-a$ to $a$ gives
\begin{eqnarray*}
\int_{-a}^{a} dx\: \phi_l^*\phi_l=-\frac{\hbar^2}{2m}\left[ \phi_l^*(\partial_x\partial_E\phi_l)-(\partial_E\phi_l)(\partial_x\phi_l^*)\right]_{-a}^{a}.
\end{eqnarray*}
By substituting $\partial_E=\frac{m}{\hbar^2 k}\partial_k$ and Eq. (\ref{eigenbasis}) we find
\begin{eqnarray*}
&&\int_{-a}^{a} dx\: \phi_l^*\phi_l\\
&&=\frac{1}{4\pi i k}(r^{l*}e^{-2ika}+r^le^{2ika})-\frac{i}{2\pi}(r^{l*}\partial_kr^l+t^*\partial_k r),
\end{eqnarray*}

\subsection{Orthogonality of the different scattering states.}
Similarly to section \ref{normalization}, we find the inner product of two different states is given by,
\begin{eqnarray*}
\langle\phi_l|\phi_r'\rangle&=&\int_{-\infty}^{\infty}dx\: \phi_l(x,k)^*\phi_r(x,k')\\
&=&\frac{1}{2\pi}\int_{-\infty}^{-a} dx (e^{-ikx}+r^{l*}e^{ikx})(t^{'}e^{-ik'x})\\
&&+\frac{1}{2\pi}\int_{a}^{\infty} dx \: t^{*}e^{-ikx}(e^{-ik'x}+r^{r'}e^{ik'x})\\
&&+\frac{1}{2\pi}\int_{-a}^{a} dx \: \phi_l^*(x,k) \phi_r(x,k')\\
&=& f_1+f_2+f_3+\int_{-a}^{a} dx \:\phi_l^*(x,k) \phi_r(x,k').
\end{eqnarray*}
In the same way as the previous paragraph, the functions $f_1$, $f_2$ and $f_3$, are defined as
\begin{eqnarray*}
f_1&\equiv&\lim_{\epsilon\rightarrow 0}\lim_{\lambda\rightarrow\infty} t^{'}\:\frac{e^{i(2k+\epsilon)\lambda}-e^{i(2k+\epsilon)a}}{2\pi i(2k+\epsilon)},\\
f_2&\equiv&\lim_{\epsilon\rightarrow 0}\lim_{\lambda\rightarrow\infty} t^{*}\:\frac{e^{-i(2k+\epsilon)\lambda}-e^{-i(2k+\epsilon)a}}{-2\pi i(2k+\epsilon)},\\
f_3&\equiv&\lim_{\epsilon\rightarrow 0}\lim_{\lambda\rightarrow\infty} (t^{'}r^{l*}+t^{*}r^{r'})\frac{e^{i\epsilon\lambda}-e^{i\epsilon a}}{2\pi i\epsilon}.
\end{eqnarray*}
For small $\epsilon$, the amplitudes, $t^{'}$ and $r^{r'}$ can be expanded around $k$ up to first order in $\epsilon$,
\begin{eqnarray*}
f_1&\simeq&\lim_{\epsilon\rightarrow 0}\lim_{\lambda\rightarrow\infty} (t+\epsilon\partial_kt)\frac{e^{i(2k+\epsilon)\lambda}-e^{i(2k+\epsilon)a}}{2\pi i (2k+\epsilon)},\\
f_2&=&\lim_{\epsilon\rightarrow 0}\lim_{\lambda\rightarrow\infty} t^{*}\:\frac{e^{-i(2k+\epsilon)\lambda}-e^{-i(2k+\epsilon)a}}{-2\pi i(2k+\epsilon)},\\
f_3&\simeq& \lim_{\epsilon\rightarrow 0}\lim_{\lambda\rightarrow\infty} \left[ (t+\epsilon\partial_kt)r^{l*}+t^{*}(r^r+\epsilon\partial_kr^r) \right]\frac{e^{i\epsilon\lambda}-e^{\epsilon a}}{2\pi i\epsilon}\\
&=&\lim_{\epsilon\rightarrow 0}\lim_{\lambda\rightarrow\infty} \left[ (t r^{l*}+t^{*}r^r)\frac{e^{i\epsilon\lambda}-e^{i\epsilon a}}{2\pi i\epsilon}\right.\\
&&\left.+(r^{l*}\partial_kt^r+t^{*}\partial_kr^r)\frac{e^{i\epsilon\lambda}-e^{\epsilon a}}{2\pi i} \right]\\
&=&\lim_{\epsilon\rightarrow 0}\lim_{\lambda\rightarrow\infty} (r^{l*}\partial_kt+t^{*}\partial_kr^r)\frac{e^{i\epsilon\lambda}-e^{\epsilon a}}{2\pi i},
\end{eqnarray*}
where we use a unitary condition of the $S$-matrix that imposes the condition $t r^{l*}+t^{*}r^r=0$ on the scattering amplitude. The very rapidly fluctuating part $e^{i(2k+\epsilon)\infty}$ and $e^{i\epsilon\infty}$ (since $\infty$ comes before $\epsilon$), averages to zero, and can be dropped. Moreover, for $\epsilon\rightarrow 0$, the functions $f_1, f_2$, and $f_3$ become
\begin{eqnarray*}
f_1&=&-t\:\frac{e^{2ika}}{4\pi ik},\\
f_2&=&t^{*}\:\frac{e^{-2ika}}{4\pi ik},\\
f_3&=&\frac{i}{2\pi}(r^{l*}\partial_k t+t^{*}\partial_k r^r),
\end{eqnarray*}
and consequently the inner product becomes
\begin{eqnarray*}
\langle\phi_l|\phi_r\rangle&=&\frac{i}{2\pi}(r^{l*}\partial_k t+t^{*}\partial_k r^r)-\frac{i}{4\pi k}(t^{*}e^{-2ika}-te^{2ika})\\
&&+\int_{-a}^{a}dx\: \phi_l^*(x,k)\phi_r(x,k).
\end{eqnarray*}
To satisfy orthogonality, we must have the condition
\begin{eqnarray}
&&\int_{-a}^{a}dx \:\phi_l^*(x,k)\phi_r(x,k)\label{orthgonality}\\
&&=-\frac{i}{2\pi}(r^{l*}\partial_k t+t^{*}\partial_k r^r)+\frac{i}{4\pi k}(t^{*}e^{-2ika}-t e^{2ika}).\nonumber
\end{eqnarray}

To prove Eq. (\ref{orthgonality}), let us consider as before the Schr\"{o}dinger equation,
\begin{eqnarray*}
H\phi&=&(T+V)\phi=E\phi,\\
(H-E)\phi&=&(T+V-E)\phi=0,
\end{eqnarray*}
and the first derivative with respect to $E$ \cite{Smith},
\begin{eqnarray*}
\partial_E\left[(H-E)\phi \right]=(H-E)\partial_E\phi-\phi=0.
\end{eqnarray*}
For a local potential and for the same coordinates eigenfunctions, as before, this equation gives
\begin{eqnarray*}
&&\phi_l^*T\left[\partial_E\phi_r\right]-\left[\partial_E \phi_r\right]\left[T\phi_l\right]^*\\
&&=\phi_l^*\left[ \phi_r-(V-E)\partial_E\phi_r\right]-\left[ \partial_E\phi_r\right]\left[T\phi_l\right]^*\\
&&=\phi_l^*\phi_r-\phi_l^*\left[V-E\right]\left[\partial_E\phi_r\right]+\left[\partial_E\phi_r\right]\left[V-E\right]\phi_l^*\\
&&=\phi_l^*\phi_r.
\end{eqnarray*}
Now we find
\begin{eqnarray*}
\phi_l^*\phi_r&=&\phi_l^*T(\partial_E\phi_r)-(\partial_E \phi_r)T\phi_l^*\\
&=&\phi_l^*\left(-\frac{\hbar^2}{2m}\partial_x^2\right)\partial_E\phi_r-(\partial_E\phi_r)\left(-\frac{\hbar^2}{2m}\partial_x^2\right)\phi_l^*\\
&=&-\frac{\hbar^2}{2m}\left[ \phi_l^*\partial_x^2(\partial_E\phi_r)-(\partial_E\phi_r)(\partial_x^2\phi_l^*) \right]\\
&=&-\frac{\hbar^2}{2m}\partial_x\left[ \phi_l^*(\partial_x\partial_E\phi_r)-(\partial_E\phi_r)(\partial_x\phi_l^*) \right].
\end{eqnarray*}
Integration from $-a$ to $a$ gives
\begin{eqnarray*}
\int_{-a}^{a} dx\: \phi_l^*\phi_r=-\frac{\hbar^2}{2m}\left[ \phi_l^*(\partial_x\partial_E\phi_r)-(\partial_E\phi_r)(\partial_x\phi_l^*)\right]_{-a}^{a}.
\end{eqnarray*}
By substituting $\partial_E=\frac{m}{\hbar^2 k}\partial_k$ and Eq. (\ref{eigenbasis}) we find
\begin{eqnarray*}
&&\int_{-a}^{a} dx\: \phi_l^*\phi_r\\
&&=-\frac{i}{2\pi}(r^{l*}\partial_k t+t^{*}\partial_kr^r)+\frac{i}{4\pi k} (t^{*}e^{-2ika}-t e^{2ika}),
\end{eqnarray*}
which is exactly same as Eq. (\ref{orthgonality}). Therefore the two eigenstates are orthogonal.



\begin{thebibliography}{99}
\bibitem{Condon} E. U. Condon, Rev. Mod. Phys. \textbf{3}, 43 (1931).
\bibitem{MacColl} L. A. MacColl, Phys. Rev. \textbf{40}, 621 (1932).

\bibitem{Buttier.Landauer} M. B\"{u}ttiker and R. Landauer, Phys. Rev. Lett. \textbf{49}, 1739 (1982).

\bibitem{Sokolovski.Baskin} D. Sokolovski and L. M. Baskin, Phys. Rev. A \textbf{36}, 4604 (1987).
\bibitem{Hauge.Stovneng} E. H. Hauge and J. A. St\"{o}vneng, Rev. Mod. Phys. \textbf{61}, 917 (1989).

\bibitem{Steinberg0}A. M. Steinberg, P. G. Kwiat, and R. Y. Chiao, Phys. Rev. Lett. \textbf{71}, 708 (1993).

\bibitem{Landauer.Martin}R. Landauer and Th. Martin, Rev. Mod. Phys. \textbf{66}, 217 (1994).

\bibitem{Winful}H. G. Winful, Phys. Rev. Lett. \textbf{91}, 260401 (2003).

\bibitem{Davies}P. C. W. Davies, Am. J. Phys. \textbf{73}, 23 (2005).

\bibitem{Aharonov.Erez}Y. Aharonov, N. Erez, and B. Reznik, J. Mod. Opt. \textbf{50}, 1139 (2003).
\bibitem{Sokolovski} D. Sokolovski, A. Z. Msezane, and V. R. Shaginyan, Phys. Rev. A \textbf{71} 064103 (2005).

\bibitem{Ordonez}G. Ordonez and N. Hatano, Phys. Rev. A \textbf{79} 042102 (2009).

\bibitem{Boyd} G. M. Gehring, A. C. Liapis, and R. W. Boyd, Phys. Rev. A \textbf{85}, 032122 (2012).

\bibitem{Pauli}W. Pauli, in \textit{Encyclopedia of Physics}, edited by S. Flugge (Springer, Berlin, 1958), Vol 5/1, p. 60.

\bibitem{Kobe1}D. H. Kobe and V. C. Aguilera-Navarro, Phys. Rev. A \textbf{50}, 933 (1994).
\bibitem{Kobe2}D. H. Kobe, H. Iwamoto, M. Goto, and V. C. Atuilera-Navarro, Phys. Rev. A \textbf{64}, 022104 (2001).

\bibitem{Werner}R. Werner, J. Math. Phys. \textbf{27}, 793 (1986).

\bibitem{Leon}J. León, J. Julve, P. Pitanga, and F. J. de Urríes, Phys. Rev. A \textbf{61}, 062101 (2000).

\bibitem{Jaworski.Wardlaw} W. Jaworski and D. M. Wardlaw, Phys. Rev. A \textbf{40}, 6210 (1989).

\bibitem{Muga2} J. G. Muga, R. Sala Mayato, and I. Egusquiza, \emph{Time in Quantum Mechanics} (Springer, Berlin, 2009), Vol.2.

\bibitem{Buttiker.Washburn} M. B\"{u}ttiker and S. Washburn, Nature \textbf{422}, 271 (2003).

\bibitem{Hartman} T. E. Hartman, J. Appl. Phys. \textbf{33}, 3427 (1962).

\bibitem{Buttiker} M. B\"{u}ttiker, Phys. Rev. B \textbf{27}, 6178 (1983).

\bibitem{Baz} A. I Baz', Sov. J. Nucl. Phys. \textbf{4}, 181 (1967); \textbf{5}, 161 (1967).

\bibitem{Rybachenko} V. F. Rybachenko, Sov. J. Nucl. Phys. \textbf{5}, 635 (1967).




\bibitem{Steinberg} A. M. Steinberg, Phys. Rev. Lett. \textbf{74}, 2405 (1995).

\bibitem{AAV} Y. Aharonov, D. Z. Albert, and L. Vaidman, Phys. Rev. Lett. \textbf{60}, 1351 (1988)
\bibitem{AV} Y. Aharonov and L. Vaidman, Phys. Rev. A \textbf{41}, 11 (1990).
\bibitem{Aharonov1}Y. Aharonov and A. Botero, Phys. Rev. A \textbf{72}, 052111 (2005).
\bibitem{Aharonov2}Y. Aharonov and L. Vaidman, Lect. Notes Phys. \textbf{734}, 399 (2008).
\bibitem{Aharonov3}Y. Aharonov, S. Popescu, and J. Tollaksen, Phys. Today \textbf{63}, 27 (2010).


\bibitem{dressel0} J. Dressel, S. Agarwal, and A. N. Jordan, Phys. Rev. Lett. \textbf{104}, 240401 (2010).
\bibitem{dressel1}J. Dressel and A. N. Jordan, Phys. Rev. A \textbf{85}, 022123 (2012).

\bibitem{dressel2}J. Dressel and A. N. Jordan, Phys. Rev. A \textbf{85}, 012107 (2012).


\bibitem{vonNeumann} J. von Neumann, \emph{Mathematische Grundlagen der Quantunmechanik} (Springer, Berlin, 1932)


\bibitem{Pollak}V. Gasparian, M. Ortu\~{n}o, J. Ruiz, E. Cuevas, and M. Pollak, Phys. Rev. B \textbf{51}, 6743 (1995).
\bibitem{Gasparian}V. Gasparian, Superlattices Microstruct. \textbf{23}, 809 (1998).

\bibitem{Dirac}P. A. M. Dirac, Proc. R. Soc. London A \textbf{114}, 243 (1927).

\bibitem{Aharonov}Y. Aharonov and L. Vaidman, J. Phys. A \textbf{24}, 2315 (1991).
\bibitem{Resch}K. J. Resch, J. S. Lundeen, and A. M. Steinberg, Phys. Lett. A \textbf{324}, 125 (2004).
\bibitem{Ravon}T. Ravon and L. Vaidman, J. Phys. A \textbf{40}, 2873 (2007).
\bibitem{Aharon}N. Aharon and L. Vaidman, Phys. Rev. A \textbf{77}, 052310 (2008).

\bibitem{Steinberg2} A. M. Steinberg, Phys. Rev. A \textbf{52}, 32 (1995).

\bibitem{Smith}F. T. Smith, Phys. Rev. \textbf{118}, 349 (1960).


















\end{thebibliography}
\end{document}